\documentclass[12pt,preprint]{aastex}

\shorttitle{Mu Observation}
\shortauthors{Thompson et al.}

\begin{document}

\title{An Observational Determination of the Proton to Electron Mass Ratio
	in the Early Universe~\footnote{
Based on observations made with ESO Telescopes at the La Silla or
Paranal Observatories under program IDs 68.A-0106 and 70.A-0017}}

\author{Rodger I. Thompson}
\affil{Steward Observatory, University of Arizona,
    Tucson, AZ 85721}
\email{rit@email.arizona.edu}

\author{Jill Bechtold}
\affil{Steward Observatory, University of Arizona,
    Tucson, AZ 85721}
\email{jbechtold@as.arizona.edu}

\author{John H. Black}
\affil{Onsala Space Observatory, Chalmers University of Technology,
SE-43992 Onsala, Sweden}
\email{John.Black@chalmers.se}

\author{Daniel Eisenstein}
\affil{Steward Observatory, University of Arizona,
    Tucson, AZ 85721}
\email{deisenstein@as.arizona.edu}

\author{Xiaohui Fan}
\affil{Steward Observatory, University of Arizona,
    Tucson, AZ 85721}
\email{fan@as.arizona.edu}

\author{Robert C. Kennicutt}
\affil{Institute of Astronomy, University of Cambridge, Cambridge CB3 OHA UK }
\email{robk@ast.cam.ac.uk}
\affil{Steward Observatory, University of Arizona,
    Tucson, AZ 85721}

\author{Carlos Martins}
\affil{Centro de Astrof\'{\i}sica, Universidade do Porto, Rua das 
Estrelas, 4150-762 Porto, Portugal}
\affil{Department of Applied Mathematics and Theoretical Physics, 
Centre for Mathematical Sciences,\\ University of Cambridge, Wilberforce 
Road, Cambridge CB3 0WA, United Kingdom}
\email{C.J.A.P.Martins@damtp.cam.ac.uk}

\author{J. Xavier Prochaska}
\affil{Lick Observatory and University of California Santa Cruz,
	Santa Cruz, CA 95064}
\email{xavier@ucolick.org}

\and

\author{Yancey L. Shirley}
\affil{Steward Observatory, University of Arizona,
    Tucson, AZ 85721}
\email{yshirley@as.arizona.edu}

\begin{abstract}

In an effort to resolve the discrepancy between two measurements of the fundamental
constant $\mu$, the proton to electron mass ratio, at early times in the universe we
reanalyze the same data used in the earlier studies.  Our analysis of the molecular 
hydrogen absorption lines in archival VLT/UVES spectra of the damped Lyman alpha 
systems in the QSOs Q0347-383 and Q0405-443 yields a combined measurement of a
$\Delta\mu/\mu$ value of $(-7 \pm 8) \times 10^{-6}$, consistent with no 
change in the value of $\mu$ over a time span of 11.5 gigayears. Here we define
$\Delta\mu$ as $(\mu_z-\mu_0)$ where $\mu_z$ is the value of $\mu$ at a redshift
of z and $\mu_0$ is the present day value. Our null result is consistent 
with the recent measurements of \citet{kin09}, $\Delta\mu/\mu = (2.6 \pm 3.0) 
\times 10^{-6}$, and inconsistent with the positive detection of a change in $\mu$
by \citet{rei06}.  Both of the previous studies and this study are based on the
same data but with differing analysis methods.  Improvements in the wavelength 
calibration over the UVES pipeline calibration is a key element in both of the
null results.  This leads to the conclusion that the fundamental constant $\mu$
is unchanged to an accuracy of $10^{-5}$ over the last $80\%$ of the age of the
universe, well into the matter dominated epoch.  This limit provides constraints
on models of dark energy that invoke rolling scalar fields and also limits the 
parameter space of Super Symmetric or string theory models of physics.  
New instruments, both planned and under construction, will provide opportunities 
to greatly improve the accuracy of these measurements.

\end{abstract}


\section{Introduction} \label{s-intro}

The values of the fundamental constants determine the nature of the physical
universe, from the size of mountains on earth to the eventual fate of the
universe as a whole.  Historically we have assumed that these constants are
invariant in space and time.  Speculation on the possibility of a time variation
of the constants was first discussed by \citet{dir37}, \citet{tel48} and \citet{gam67}.  In
very rare cases, such as the Oklo mine \citep{dam96}, there exists a terrestrial 
laboratory to test for time varying constants. It has been known for over thirty 
years \citep{thm75} that Damped Lyman Alpha systems (DLAs) \citep{wol05} offer 
the opportunity to measure the value of the fundamental
constant $\mu$, the proton to electron mass ratio\footnote{Although the 
literature is approximately equally divided in usage we designate $\mu$ as the
proton to electron mass ratio rather than the inverse to be consistent with the
other recent astronomical determinations of $\mu$ discussed here}, at early
times in the universe. The opportunity stems from the direct dependence of the rotational
energy of molecules on $\mu$ and the square root dependency on $\mu$ of the
vibrational energy relative to the electronic energy\footnote{See \cite{shu91}
chapter 28 for an alternative derivation of the dependence}.  Each absorption 
line has a unique shift for a change in $\mu$ that depends on the vibrational 
and rotational quantum numbers of the upper and lower energy states.  At the time of 
\citet{thm75}, however, the observational capabilities of astronomical spectroscopy and 
the accuracy of molecular hydrogen laboratory spectroscopy allowed only very crude 
determinations of $\mu$ at relatively modest look back times.  The high line
density of atomic hydrogen lines in DLAs and the rarity of DLAs with measurable
amounts of molecular hydrogen further complicated progress.

\citet{fol88} and \citet{cow95} made early measurements of $\mu$ at significant 
look back times and found no change to accuracies of $\Delta\mu/\mu \leq 
2 \times 10^{-4}$ and $7 \times 10^{-4}$ in the spectrum of PKS 0528-250 at a redshift of 2.811.  
At the same time calculations of the expected shifts were made by \citet{var93} 
who developed a method of sensitivity constants for each line that will be discussed 
later in this work. An additional constraint of $2 \times 10^{-4}$ was obtained on the 
same object by \citet{pot98}. An excellent review of studies relevant to a determination 
of the time history of $\mu$ and other fundamental constants is given in \citet{uza03}. 

Three advances now provide the opportunity to measure $\mu$ at large
look back times and at accuracies that are starting to
impact other areas of physics such as dark energy and string theory.
The first advance is the construction of large telescopes such as the Keck
telescopes, the Very Large Telescopes (VLT) and now the Large Binocular
Telescope (LBT).  A second advance is the installation of stable,
high resolution and sensitive spectrometers such as HIRES \citep{vog94}
at Keck and UVES \citep{dek00} 
at the VLT.  The third key advance is the measurement of the 
wavelengths of the H$_2$ Lyman and Werner electronic transitions to
accuracies of a few parts in $10^8$ \citep{uba07}.  In addition 
\citet{uba07} has recalculated the sensitivity constants, taking into
account both adiabatic and nonadiabatic perturbations, to provide
an invaluable set of wavelengths at the present day value of $\mu$ and
wavelength sensitivities to $\mu$ for the evaluation of the astronomical
observations.

The most recent efforts to measure $\mu$ at high redshifts have centered on
the Ultraviolet and Visible Echelle Spectrometer (UVES) on the VLT.  The 
spectra of two quasars observed in January of 2002 (Q0347-383) and January
of 2003 (Q0405-443) contain H$_2$ absorption lines at redshfits of 3.0249 
and 2.5947.  The first observations of Q0347-383 were commissioning observations
carried out in 1999 and described by \citet{dod01}.  \citet{iva02} used this
data along with a UVES spectrum of Q 1232+082 to investigate possible changes in 
$\mu$. They found two results, $\Delta\mu/\mu = (5.7 \pm 3.8) \times 10^{-5}$
and $\Delta\mu/\mu = (12.5 \pm 4.5) \times 10^{-5}$ at the 3$\sigma$ level for two
different sets of Thorium Argon wavelength lists.  A subsequent analysis by 
\citet{lev02} using just the Q0347-383 spectra found a result of $ -1.5 
\times 10^{-5} \leq \Delta\mu/\mu \leq 5.7 \times 10^{-5}$. A later reanalysis
of the Q0347-382 data by \citet{iva03} produced a limit at a confidence
level of $95\%$ of $\mid \Delta\mu/\mu \mid < 8 \times 10^{-5}$. 
\citet{uba04} combined the line lists of \citet{iva02} and \citet{lev02} 
and found that $\Delta\mu/\mu = (-0.5 \pm 3.6) \times 10^{-5}$ at the 2$\sigma$ level.

The 2002 and 2003 UVES VLT observations of Q0347-383 and Q0405-443 \citep{iva05}
had higher signal to noise than the 1999 observations.  Using new laser
determined H$_2$ wavelengths from \citet{phi04} and the UVES pipeline reduction
of the spectra they found $\Delta\mu/\mu =(1.64 \pm 0.74) \times 10^{-5}$. 
\citet{rei06} subsequently utilized a new set of laser determined H$_2$
wavelengths and the pipeline data to find a change in $\mu$ of $\Delta\mu/\mu
= (2.4 \pm 0.6) \times 10^{-5}$.  \citet{uba07} details the determination of 
the H$_2$ parameters and gives a more complete list of laser determined
wavelengths that slightly alters the result to $\Delta\mu/\mu = (2.45 \pm 0.59) 
\times 10^{-5}$.  The \citet{uba07} H$_2$ parameters essentially
remove the properties of H$_2$ from the error budget leaving the data 
reduction and signal to noise of the observed spectrum as the primary
error contributors.  The \citet{rei06} result for Q0347-383 was examined 
by \citet{wen08} who concluded that the data were consistent with 
$-0.7 \times 10^{-5} \leq \Delta\mu/\mu \leq 4.9 \times 10^{-5}$ at the 
$95\%$ confidence level. \citet{kin09} have taken the same data set as
\citet{rei06} with the addition of spectra of Q0528-250 and found a value
of $\Delta\mu/\mu = (2.6 \pm 3.0) \times 10^{-6}$ for the combined data set.  
A key element in their analysis is an improved wavelength calibration
as described in \citet{mur08}. 
At this time there are two analyzes of the same data that lead to two 
different conclusions. Our independent analysis of the same data concludes 
that there is no evidence for a change in $\mu$, consistent with the results of 
\citet{kin09}.

For completeness we consider radio frequency measurements of $\mu$ that
are more precise but at significantly lower redshift.  Although the
wavelength determinations are more precise, transitions in different
molecules must be compared to provide information on any change in $\mu$.
Recently \citet{fla07} have looked for variations in $\mu$ using the radio
emission lines of ammonia and carbon monoxide.  They take advantage of
the high sensitivity of the inversion spectrum of ammonia to changes in
$\mu$ with $ \Delta\mu/\mu = 0.289 \frac{z_{inv} - z_{rot}}{1 + z_0}$
where $z_{inv}$ is the redshift of the inversion lines of ammonia,
$z_{rot}$ is the redshift of the rotational lines of CO and $z_0$ is 
the cosmological redshift of the galaxy. For the galaxy B0218+357 at
a redshift of 0.68470 they find $\Delta\mu/\mu = (0.6 \pm 1.9) \times
10^{-6}$. \citet{mur08b} have improved this result to 
$\Delta\mu/\mu \le 0.18 \times10^{-6}$.  This result is at 
relatively low redshift and it depends on ammonia and carbon monoxide 
having identical kinetic velocities in the molecular clouds. This is 
probably unlikely since, unlike the ubiquitous CO molecule, NH$_3$ is
concentrated in the colder denser cores of molecular clouds.  The fact 
that it is a null
result, however, adds credence to the result since an offset in kinetic
velocity would have to accurately match any change in $\mu$ to produce
a null result. The result is also for a relatively low redshift, placing it 
within the current dark energy dominated epoch of the universe. Some
dark energy theories predict that the fundamental constants only roll
during the matter dominated epoch and freeze out at their present
values once dark energy becomes dominant around a redshift of 1, 
(eg. \citet{bar02}). Table~\ref{tab-comp} provides a summary of the 
astronomical determinations of $\mu$.

In our own galaxy \citet{lev08} have reported variations in $\mu$ based
on the same ammonia transition along different lines of sight. In this
case the variation is relative to the CCS molecule and is manifested
by a general positive velocity offset between the ammonia and CCS
emission lines.  Their result gives $\Delta\mu/\mu = (4-14) \times 10^{-8}$.
Slight errors in the line frequencies could mimic such a result.

Laboratory experiments have set limits on the present day rate of change
of $\mu$.  Even though their time base is brief by cosmological 
standards, their wavelength accuracy is far better than can be 
achieved in astronomical observations.  The current best laboratory
limits appear to be the results of \citet{bla08} which give a result
of $\dot{\mu}/ \mu = (1.6 \pm 1.7) \times 10^{-15}/yr$. To put this
in perspective if the rate of change is constant at $10^{-15}/yr$ then 
the change at the 11 gigayear look back time of Q0347-383 would be
$1.1 \times 10^{-5}$, similar to the astronomical results given
in this work.  There is no real expectation that the rate of change 
would be constant so both the astronomical and laboratory results work
in concert to constrain possible physical models that predict changes
in time of the values of the fundamental constants.  The results of
\citet{bla08} depend on the Schmidt model for the nuclear magnetic
moment and therefore may be deemed as model dependent.  A laboratory
result that is independent of the Schmidt model is given by \citet{she08}
who give $\dot{\mu}/ \mu = (-3.8\pm 5.6) \times 10^{-14}/yr$.  
Other limits on the
present rate of variation in $\mu$ based on the Weak Equivalence Principle 
and various theories of particle physics are discussed by \citet{den08}.

The remainder of the paper addresses the measurement of $\mu$ in the
spectra of Q0347-383 and Q0405-443. The wavelength calibration and
data reduction to produce the spectra used in this work will only
be summarized since it is discussed in detail in \citet{thm09}. That
separate publication is intended to give a full description of the
data analysis in order to allow the reader to concentrate on the 
measurement of $\mu$ described here without a lengthy data reduction
description at the beginning.  In this paper we bring those analysis
methods to bear in an effort to discriminate between the positive
and null results for a variation in $\mu$.

\section{Observations} \label{s-obs}

The observations of Q0347-383 and Q0405-443 with UVES on VLT occurred during the
nights of January 7-9 2002 for Q0347-383 and January 4-6 2003 for 
Q0405-443\footnote{Based on observations made with ESO telescopes at the Paranal 
Observatory under program IDs 68.A-0106 and 70.A-0017}. 
The emission line redshifts for these QSOs are 3.22 and 3.02 respectively
\citep{iva05}. 
The data were retrieved from the VLT archive along with the MIDAS based UVES 
pipeline reduction procedures.  On each of the nights three separate spectra of
the QSO were taken with accompanying long slit calibration lamp integrations at 
the same grating setting. The slit width and length for both object and 
calibration line observations are 0.8 and 6.6 arc seconds.  The grating angle for the
Q0347-383 observations had a central wavelength of 4300 \AA\ and for Q0405-443,
3900 \AA\ . The images are 2x2 pixel binned on chip with a size of 1024 by 1500 
binned pixels.  A single pixel is 15 microns in size and 0.22 arc seconds on
the sky. In the following the word pixel refers to the 2x2 binned pixels (0.44
arc seconds) in the images obtained from the archive.  At 4000 \AA\ a pixel
is approximately 0.0416 \AA\ which is about 3 km/sec.
Both the calibration and object images are binned identically. Exposure times
and other observational parameters are given in Tables~\ref{tab-obs7} and 
\ref{tab-obs5} and are described in \citet{iva05}.

There are differences in the way the long slit calibration spectra were taken
between the two objects.  In the case of Q0405-443 there was a long slit 
calibration spectrum taken immediately after the object spectrum in all but
one case.  The time tags of the grating position encoder readouts for the 
object and calibration spectra are identical as are the values of the 
grating position encoder readings.  This indicates that there was
no adjustment of the grating position between the paired object and calibration
spectra. The exception is the night of January 5, 2003 where there is no long
slit calibration spectrum for the second object observation. 

For Q0347-383 two long slit calibration spectra were taken in between the three
object spectra for each night.  The encoder readings indicate that there were 
no grating resets performed between the object spectrum and the calibration 
spectrum for the first two pairs of object and calibration observations for 
each night.  The time tags, however, on the third night of January 10, 2002
pairs the long slit calibration spectrum with the third object spectrum.  The
proper pairings of observations are important in calculating the shifts needed
to accurately combine the observations as is discussed in \S~\ref{ss-wvc}.

\section{Data Reduction} \label{s-dr}

The spectra described by \citet{iva05} and used by \citet{rei06}
were produced by the standard UVES pipeline.  The pipeline produces excellent 
spectra for most observations, however, \citet{mur08} points out that the Thorium
Argon line list used in the wavelength calibration may not be accurate enough
for the precise determination of fundamental constants.  We reached similar 
conclusions as discussed in \S~\ref{ss-wvc}.  The final output of the UVES
pipeline is an interpolated spectrum with equal wavelength intervals as opposed
to an intensity and wavelength on a pixel by pixel basis.  In what follows we 
only use images and spectra on a pixel by pixel basis.  The wavelength
calibration and the production of the spectra used in this study are
described in detail in \citet{thm09}.  The descriptions given here are
short summaries of the methods.

\subsection{Wavelength Calibration} \label{ss-wvc}

Independently of \citet{mur08} we became aware that the standard Th/Ar
line list used in the UVES pipeline analysis provides good wavelengths
for most studies but is the primary limiting factor in obtaining the 
accuracy required for a determination of $\mu$ at the $10^{-5}$ level.
In fact only about 1/4 of the lines are free of blending and other
problems.  We then recalibrated the wavelength solutions using the long
slit calibration line spectra taken during the observations of the two
QSOs.  This is described in detail in \citet{thm09} which is intended
to serve as the record of the wavelength calibration used in this study
and therefore will not be repeated here.  The new wavelength calibration
is the primary reason for a null result in this study.  It should be 
noted that this recalibration
differs from the recalibration used in \citet{kin09} in two ways. 
First, this calibration is based on the calibration spectra taken during
the observation of the analyzed spectra.  Second the calibration is done
order by order.  This results in some lines being declared good in one
order but unusable in another order where they fall in low signal to
noise areas.

The wavelength calibration described in \citet{thm09} tracks the shifts
in the wavelength positions between observations and between different
observing nights.  The final wavelength calibration is relative to a
master solution which is set at a single position determined by a master
long slit calibration lamp image that is the median of all of the
calibration lamp images shifted to the position of the first calibration 
lamp image.  The shifts are small, a few hundredths of a pixel width,
but important in this study. The shifts are carried out by cubic 
interpolation and are rigid.  The wavelength solution for each order
of the master long slit calibration image is a six term Legendre
polynomial whose coefficients are different for each order.  The 
wavelength solution for the object spectra will differ from the
master solution first due to small shifts in the actual grating
position from the position appropriate to the master solution and
due to the motion of the observatory about the barycenter of the
earth-sun system.  These are corrected for in the production of
the individual spectra.

\subsection{Spectrum Production} \label{ss-sp}

The order by order final spectrum for each object is a three dimensional 
array of dimensions [np,nord,6] where np is the number of pixels in
the dispersion direction (1500), nord is the number of orders and the
six last dimensions are flux, wavelength, variance, continuum, fit, and
the fit convolved with the instrumental profile for each pixel in the 
spectrum.  In this case the fit is the continuum minus the best fit to
the H$_2$ lines at their natural line width.  This is what the spectrum
would look like if the instrument profile was infinitely narrow.  
The first 2 are derived from
the object and calibration spectra and calculated for each spectrum.
The last 4 are calculated after the spectra are combined into 
a single spectrum but could be, in principle, calculated for the
individual spectra.  The observational parameters for Q0347-383 and
Q0405-443 are listed in Tables~\ref{tab-obs7} and \ref{tab-obs5}
respectively.

\subsubsection{Flux} \label{sss-flx}

At this stage there are 9 spectra for each of the 2 objects.  In
each order of the spectra the flux is distributed over several pixels
in the cross dispersion direction. We tested several optimal extraction 
methods for combining the flux in the cross dispersion direction 
into a single value.  These tests indicated that the UVES MIDAS V2.2.0 
pipeline extraction did as good or better job of combining the flux 
than any of the methods we tested.  We therefore used this intermediate
product of the pipeline to produce the 9 flux versus pixel spectra
for the 2 objects. These are not the interpolated to constant delta 
wavelength values spectra that are the final product of the pipeline. The next 
step is to assign a proper wavelength designation
to each of the pixel positions in each order.  Our goal is a proper
vacuum wavelength at rest with respect to the barycenter of the earth
sun system.  Again note that our reference to pixels is to the 2x2 pixel
binned output available in the archive.

\subsubsection{Wavelength} \label{sss-wav}

As mentioned in \S~\ref{ss-wvc} the true wavelengths of the pixels are slightly 
different for each spectrum due to small shifts in the spectrometer configuration 
and different barycentric velocities.  The wavelengths are calculated
by first shifting the master wavelength solution by the amount calculated
from the associated long slit calibration spectrum and then correcting
the wavelengths for barycentric velocity.  The associated long slit calibration
line observations are given in \citet{thm09}.

The wavelength shift due to small differences in the grating position 
is calculated from the shifts found during the wavelength calibration. As
described in \S~\ref{s-obs} care was taken to not reset the grating
position between a long slit calibration observation and its associated
object exposures.  For these observations the shift of the object spectrum 
wavelength position is identical to the shift calculated for the associated
calibration line spectrum given in \citet{thm09}.  There are, however, two 
object observations where it appears that the grating position was reset 
without an associated long slit calibration line exposure. They are the 
observations in Table~\ref{tab-obs7} for Q0347-383 that end in 109\_b and 981\_b.  
For these observations our only recourse was to take the shift as the average 
between the shifts immediately preceding and
immediately after the observations.  The correlation between the encoder
readouts and the shifts calculated during the wavelength calibration
did not appear to be accurate enough to be used as a direct 
indicator of the amount of shift.  Once the shift is determined
the master wavelength solution is interpolated to account for the
shift in pixel position between the associated calibration line
image and the master image.  The wavelengths associated with the
flux in the pixels now have the correct observed vacuum wavelength
as observed but must still be corrected for the barycentric
velocity.  Note that the handling of the shift values is different
from the description given in \citet{thm09}.  At  that time is was
not known that there was a procedure of reseting the grating position
between observations.  It does not affect the master wavelength
calibration since the shifts were calculated directly from the
calibration observations, but it does matter for the object spectra.

The component along the direction to the object of the barycentric velocity
of the observatory was calculated using the date and time of the midpoint
of the integration. This velocity is due to the earth's orbit and rotation 
relative to the barycenter of the earth-sun system.  The wavelength scale
was then corrected for this motion so that the final wavelengths are
vacuum wavelengths as observed in a reference frame at rest relative to
the barycenter.  This is slightly different than the heliocentric 
wavelengths used in \citet{iva02}.

\subsection{Co-addition of the Spectra}

At this point the individual spectra have the flux, wavelength
and noise values populated in their respective arrays.  The
wavelengths, however, are slightly different for each spectrum
due to the grating shifts and different barycentric velocities.
Accurate co-addition of the spectra requires that they all be
on the same wavelength scale.  We choose to shift all of the spectra
to the wavelength scale of the master wavelength solution for the
grating angle setting as described in \citet{thm09}.

\subsubsection{Shift to Common Wavelength Scale}

The spectrum shift was accomplished by interpolation of the flux 
from the wavelength scale of the spectrum to the master wavelength 
scale of the master solution using the IDL\footnote{IDL stands 
for Interactive Data Language registered by ITT Visual Information 
Solutions} code procedure \emph{INTERPOL} in double precision mode. In the 
following we will write IDL provided procedure names in capital italic 
letters and procedures written by the authors using IDL code in lower 
case italics.  \emph{INTERPOL} uses linear interpolation which
is appropriate for this case since the shifts are only a few
hundredths of a pixel.  For cases with significant fractions of a
pixel other interpolation methods may be more appropriate.  At the
end of this procedure all spectra are on a common wavelength scale,
the master wavelength solution for each order.

\subsubsection{Addition of the Spectra}

All of the observed spectra for each of the two objects were combined 
to produce the final two spectra for analysis.  The excellent and
uniform observing conditions at the VLT produced a suite of individual
spectra with remarkably similar signal to noise characteristics.  In
other words the weight of each of the spectra were essentially indistinguishable
from each other.  For this reason we simply produced two spectra for
each object, one that is the mean of all the flux values at a given
wavelength and the second which is the median of the flux values.  Again
the differences between these two spectra were minimal.  The median
spectrum was judged to have slightly better signal to noise in both
objects and is the spectrum that is used in the analysis.

\subsubsection {The Variance} \label{sss-var}

The UVES pipeline calculates the variance for each of the spectrum fluxes
but the documentation is not clear on the exact method of calculation.  
The variance is an important quantity in calculating the $\chi^2$ values for the 
wavelength and density fits so uncertainty in how it was calculated is 
worrisome.  The variance is therefore calculated explicitly form the 9
spectra normalized to a common total flux value.  The normalizations varied
between 0.8 and 1.2 for the 9 spectra that were combined to make the final
spectrum for each object.  The normalization, therefore, does not have a
large effect on the calculated variance.

\subsection{H$_2$ Line Parameters}

A primary component of this study is the use of accurate molecular
data provided by several recent studies of the H$_2$ molecule 
(\citet{uba07}, \citet{iva08} and the references therein).  The 
data from these references include the vacuum wavelengths and 
calculated sensitivity factors K$_i$ where the index i indicates
the line.  The sensitivity factor to a variation in $\mu$ is 
different for each line and is defined as

\begin{equation} \label{eq-k}
K_i = \frac{d \ln \lambda_i}{d \ln \mu} = \frac{\mu}{\lambda_i}\frac{d \lambda_i}
{d \mu}
\end{equation}

The precise vacuum wavelengths from these references have average 
errors on the order of $5 \times 10^{-6}$ \AA\ which produce a 
negligible contribution to the errors in determining the redshift of each 
line.  The oscillator strengths for each transition were calculated using 
the Einstein A coefficients from \citet{abg93a} for the Lyman transitions 
and \citet{abg93b} for the Werner transitions.  

\subsection {Output Products}

The calculation maintains two primary output products.  The first is a six
component spectrum of each order.  The six components are the double precision
wavelength, the flux, the standard deviation of the flux, the continuum fit,
the line fit and the line fit convolved with the instrument profile.  The 
second output is an IDL structure array which contains the line information.
An IDL structure is a multi-format data set that can contain text, integer,
floating point and arrays of any of these formats.  There is a structure
for each line which contains the molecular data such as oscillator strength
and vacuum rest wavelength as well as the calculated data from the fit such as
redshift and density.  Information on whether the fit for the line converged
is also in the structure.

\section{Analysis of the H$_2$ Lines} \label{s-h2}

The purpose of this analysis is to determine as accurately as possible
the true vacuum wavelength of the observed H$_2$ absorption lines in
the spectra of Q0347-383 and Q0405-443 produced by the procedures
discussed in \S~\ref{s-dr} and \citet{thm09}.  These wavelengths are
then used in \S~\ref{s-dm} to measure the value of $\mu$ at the epoch
represented by the redshift of the DLA absorption line system. The 
procedures described in the following are procedures written in 
IDL.

\subsection{Establishing the Continuum} \label{ss-cont}

Our definition of the continuum in this section is not the true continuum
of the quasar spectrum but rather the true spectrum without the H$_2$
absorption lines.  This is the canvas that the H$_2$ spectrum is painted
on.  In an analogy to preparing a canvas by sizing it we refer to this as
sizing the continuum.  The procedure starts with a first guess of the 
column densities of the first 5 rotational levels of the H$_2$ electronic
and vibrational ground state and calculates the expected line width, defined
as all regions less than $90\%$ of the continuum, using a Voigt function, the
oscillator strength of the transition, a temperature of 350K and the redshift
of the H$_2$ absorption line system.  The kinetic temperature of 350K is 
equivalent to Doppler parameter of 1.7 km s$^{-1}$. Initially the redshift 
was taken from previous studies but subsequent iterations used the best 
redshift from the previous iteration. There is only one velocity component 
for Q0347-383 and only the strongest component of the two velocity components 
in Q0405-443 was used. The natural line shape is convolved with the UVES 
instrumental profile to produce the observed line profile.  This procedure
simply defines the spectral region that must the replaced with the continuum
estimate. The spectral region inside the line width is then replaced with 
the expected continuum calculated by interpolating the spectrum on either 
side of the line which has undergone a 5 pixel smoothing. It is important to
note that as mentioned at the beginning of the section this is a local
continuum fit which represents what the spectrum would be if there were
no molecular hydrogen lines, not the true continuum flux.  Since the fit
is local there is an independent continuum fit for each line. The typical 
line spans two to three pixels. The automated continuum calculation gives 
a rough, first order fit to the sized continuum. At this point a hard copy 
plot of the spectrum and continuum is produced and examined.  

\begin{figure} 
\plotone{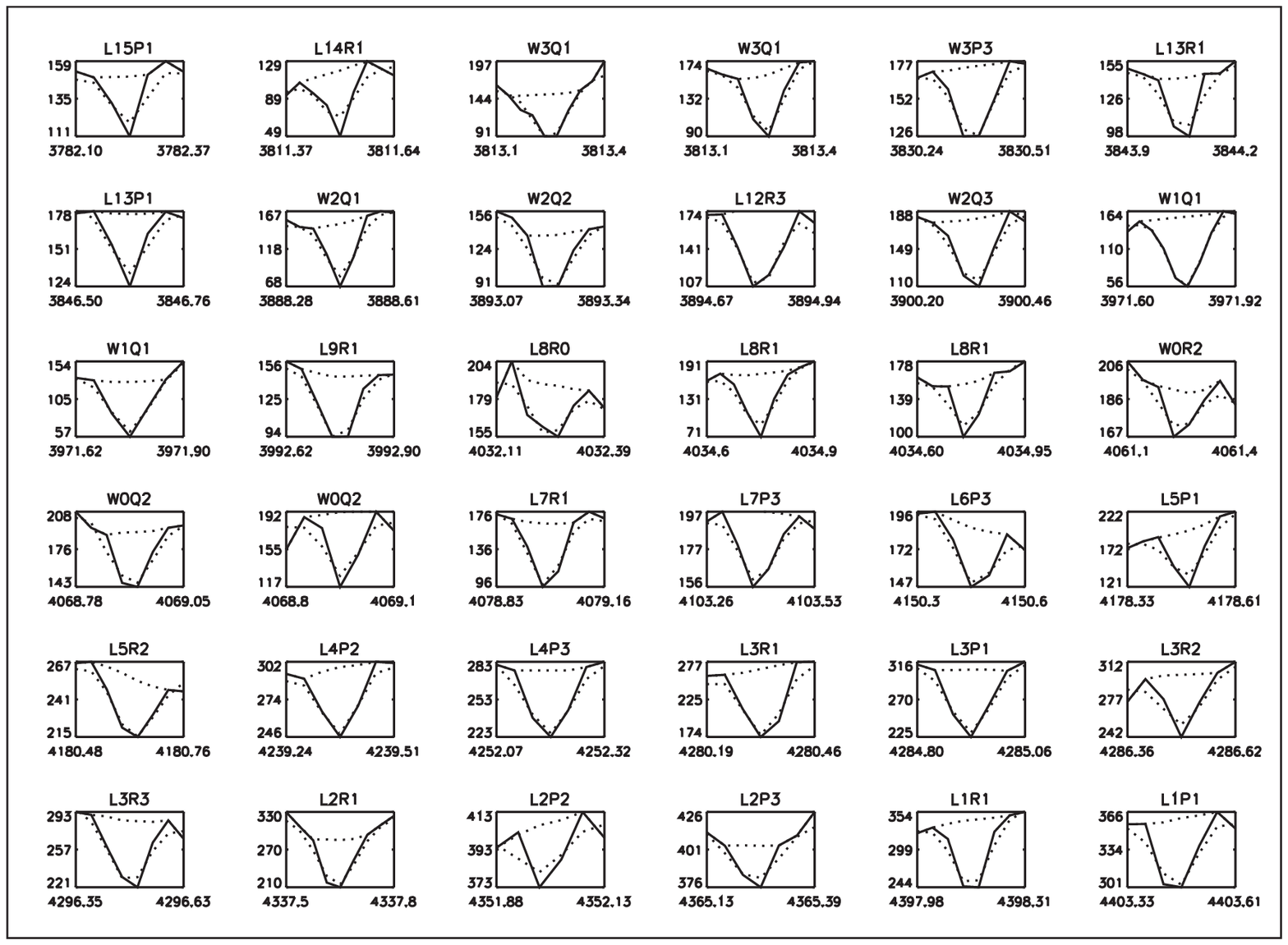}
\caption{The spectrum of Q0347-383 in the regions of the H$_2$
lines used in this analysis is shown by the solid line.  The 
adopted continuum and fits convolved with the instrument profile
are shown by the dotted lines.  Repeated transitions are the 
same H$_2$ line but in different orders.  Transitions are labeled 
in the same manner as in Tables~\ref{tab-d47} and \ref{tab-d45}.
The intensities are give in ADUs per second.
\label{fig-m4}}
\end{figure}

\begin{figure} 
\plotone{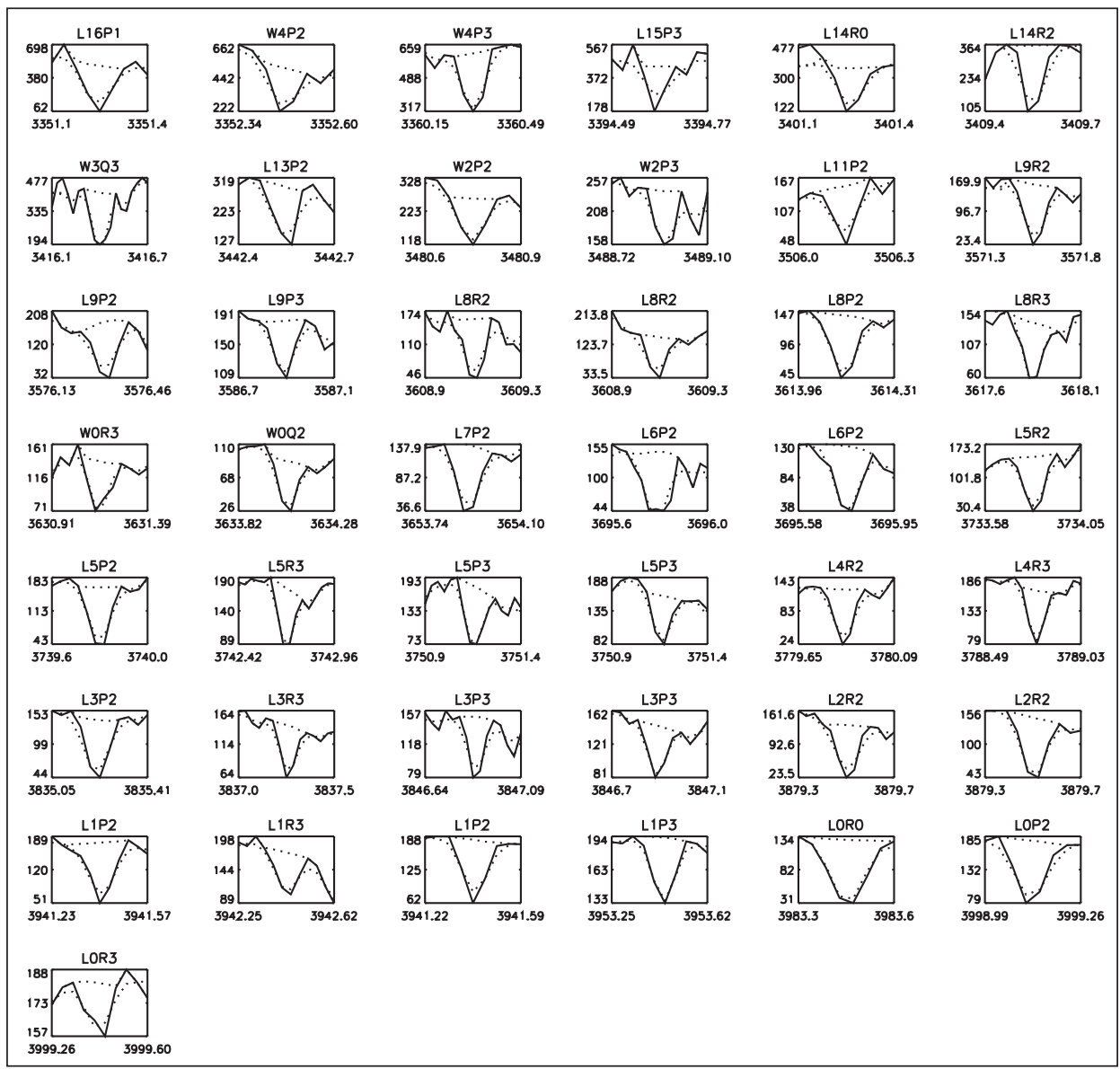}
\caption{The spectrum of Q0405-443 in the regions of the H$_2$
lines used in this analysis is shown by the solid line.  The 
adopted continuum and fits convolved with the instrument profile
are shown by the dotted lines.  Repeated transitions are the 
same H$_2$ line but in different orders.  Transitions are labeled 
in the same manner as in Tables~\ref{tab-d47} and \ref{tab-d45}.
\label{fig-m3}}
\end{figure}

The continuum fit of the selected lines is then interactively refined in an IDL
based procedure that displays the fit for each line and allows the user to
adjust the fit interactively.  The complex spectrum of the Lyman Alpha forest
insures that most of the continuum fits must be adjusted.  Since the goal of 
this study is the most precise possible wavelength fit, rather than a column
density fit, the continuum is usually adjusted preferentially to lower 
values that emphasize the line center as opposed to the lower signal to noise 
wings of the line. Tests with various continuum fits indicated that adjustment 
of the continuum to higher levels produced significant errors in wavelength 
for some lines but that adjustment to a reasonable range lower levels did not
produce wavelength changes larger than the wavelength errors determined in 
\S~\ref{sss-chi}.  Apparently the larger number
of pixels in the high continuum cases allowed noise in the wings of the line
to have a greater influence on the fit than with the low continuum case. The
line shape convolved with the instrumental profile is about twice as wide as the
natural line shape for unsaturated lines. A mosaic of the spectral regions of 
all of the lines used in the analysis along with the continuum and line fits 
are shown in Figures~\ref{fig-m4} and \ref{fig-m3}. These spectra are displayed
to allow the reader to judge the quality of the continuum and line fits. The 
spectra at this point are ready for the $\Delta\mu/\mu$ determination.

\section{Determination of the $\Delta\mu/\mu$ Value} \label{s-dm}

\subsection{Selection of Suitable H$_2$ Lines for the Measurement of $\mu$}

Most of the H$_2$ lines are unusable due to the Lyman Alpha forest.  
Appropriate lines are picked at this time based on freedom from interference 
by other lines and signal to noise.  A basic selection rule is lines that have 
greater than $50\%$ asymmetry between the height of their short wavelength and 
long wavelength shoulders are rejected.  This limits the number of lines used 
that lie on the shoulders of other lines.  Lines that have asymmetric profiles, 
indicating a blend of two lines, are also rejected. Lines that have profiles 
broader by $50\%$ than expected from a single line are similarly
rejected.  Finally lines that do not converge to a stable redshift value
in the following analysis are not used to determine the $\mu$ value. During 
the course of the analysis slightly different selection rules were applied.   
More lenient rules led to larger errors as did more stringent rules that 
reduced the number of lines in the analysis.  In no case, however, did the 
results exceed a 2$\sigma$ excursion from a null result.  The
lists of lines used in this analysis are given in Tables~\ref{tab-d47} 
and~\ref{tab-d45}.  These line lists do not directly correspond to the lines
used by \citet{rei06} who did not publish their selection criteria but were
guided by lines selected by \citet{iva05}. \citet{kin09}
do not list the lines that they used but the text indicates that they were not
the same as those use by \citet{rei06}.  They did, however, conduct an analysis
using the same lines as \citet{rei06} and obtained a result of $\Delta\mu/\mu = 
(12.0 \pm 14.0) \times 10^{-6}$.  As in the analysis of \citet{rei06} only the
stronger of the double system of H$_2$ lines in Q0405-443 were used.  The two
systems are separated by 13 km/sec which is a separation of roughly 4 of the 
double binned pixels.

\subsection{Fitting the Lines} \label{s-fl}

The line fitting procedures are IDL based double precision procedures 
written and developed by authors.  Each line is fit individually rather
than calculating a complete synthetic spectrum for several reasons.  
The first is that most of the large number of molecular hydrogen lines
are unusable due to blends with other lines or due to complete obliteration
by the Lyman $\alpha$ forest.  They would simply contribute noise to
the fit.  Second we are looking for shifts away from the expected
wavelengths that a global fit would wash out.  Finally we allow the 
column density to be an independent parameter for each line and use 
anomalous densities to find lines that are blended with other lines.  

The individual selected H$_2$ lines 
are fit iteratively with alternate adjustments of the wavelength and the column
density.  The fit function is a Voigt function calculated with the IDL
function \emph{VOIGT} that is convolved with the instrument profile. The IDL
\emph{VOIGT} function is calculated with double precision parameters and returns 
a double precision result.  The instrument profile is represented by a 
Gaussian of halfwidth 0.037014 \AA\ at 3900 \AA\ digitized in units of 
0.001 \AA\ \citep{car05}. The halfwidth is adjusted at other wavelengths 
to be directly proportional to the wavelength.  Changes in the width of 
the Gaussian by plus or minus $10\%$ changed the derived column density 
but had no effect on the derived wavelength within the wavelength errors 
attributable to signal to noise.  The kinetic and excitation temperature 
of the gas is not varied but held at 350K for both objects for the initial
fitting.  During the iteration of the fits we did not require all lines
with the same lower state to have the same column density. This is similar to 
letting the excitation temperature vary from line to line. The kinetic 
temperature is held fixed at 350K. Changes in the kinetic temperature 
by $\pm 100$ K did not alter the derived wavelengths within the $1 \sigma$ 
bounds.

The fit is started with an initial guess at the column density for each ground state
rotational level and an initial guess at the redshift.  After a few runs
these initial guesses were refined to produce a better starting solution.  The
fit procedure starts with the wavelength adjustment followed by a column density 
adjustment.  This procedure is iterated 6 times.  Lines that have not converged
after six iterations are then rerun with another 6 tries at convergence.  Any
lines that have not converged in both column density and wavelength after the two
iterations are not used in the analysis.  Convergence is declared when two
tries in sequence return the same column density and wavelength values.

\subsubsection{Wavelength Iteration}

The first of the 6 tries in each iteration starts with a sweep of the wavelength
in 200 $10^{-5}$ \AA\ steps on either side of the starting wavelength.  The starting
wavelength is either the wavelength from the initial guess at the redshift for the
first iteration or the best wavelength from the previous try in the second 
and subsequent iterations.  At each of the 200 wavelengths on either side of 
the initial wavelength the line  fit is calculated as a Voigt function superimposed 
upon the continuum spectrum as discussed in \S~\ref{ss-cont} and \S~\ref{s-fl}.  
The calculated spectrum is then convolved with the instrument profile as 
described in \S~\ref{s-fl}.  A $\chi^2$ value for the difference between the fit
and the observed flux is then calculated for
all spectral points in the line that are deeper than $95\%$ of the continuum.  
After the sweep over the 400 wavelength positions the best wavelength is taken
to be the wavelength with the minimum $\chi^2$ value.  If the best wavelength is
not at either extremum of the wavelength sweep the number of test wavelengths is 
reduced by a percentage that is proportional to the distance of the best wavelength
from the extremum.  The minimum number of test wavelengths is 10.  Subsequent 
tries at the wavelength fit are all centered at the best wavelength from the 
previous try.

\subsubsection {Column Density Iteration}

After each try at the wavelength fit there is an adjustment of the column density.
The wavelength is fixed at the best wavelength from the last wavelength iteration. 
The column density is varied over 200 values ranging from $1\%$ of the 
initial density to twice the initial density in $1\%$ increments.  The initial
column density is either the initial guess column density or the density found
in previous column density iteration. A $\chi^2$ value is then calculated for 
all column density points in the same way as the wavelength iteration.  The best 
column density is taken as the density with the minimum $\chi^2$ value. 
Although the column density range is asymmetric between the high and low ends 
it usually converges in the first 2 to 3 iterations.

\subsubsection {The $\chi^2$ Values} \label{sss-chi}

\begin{figure} 
\plotone{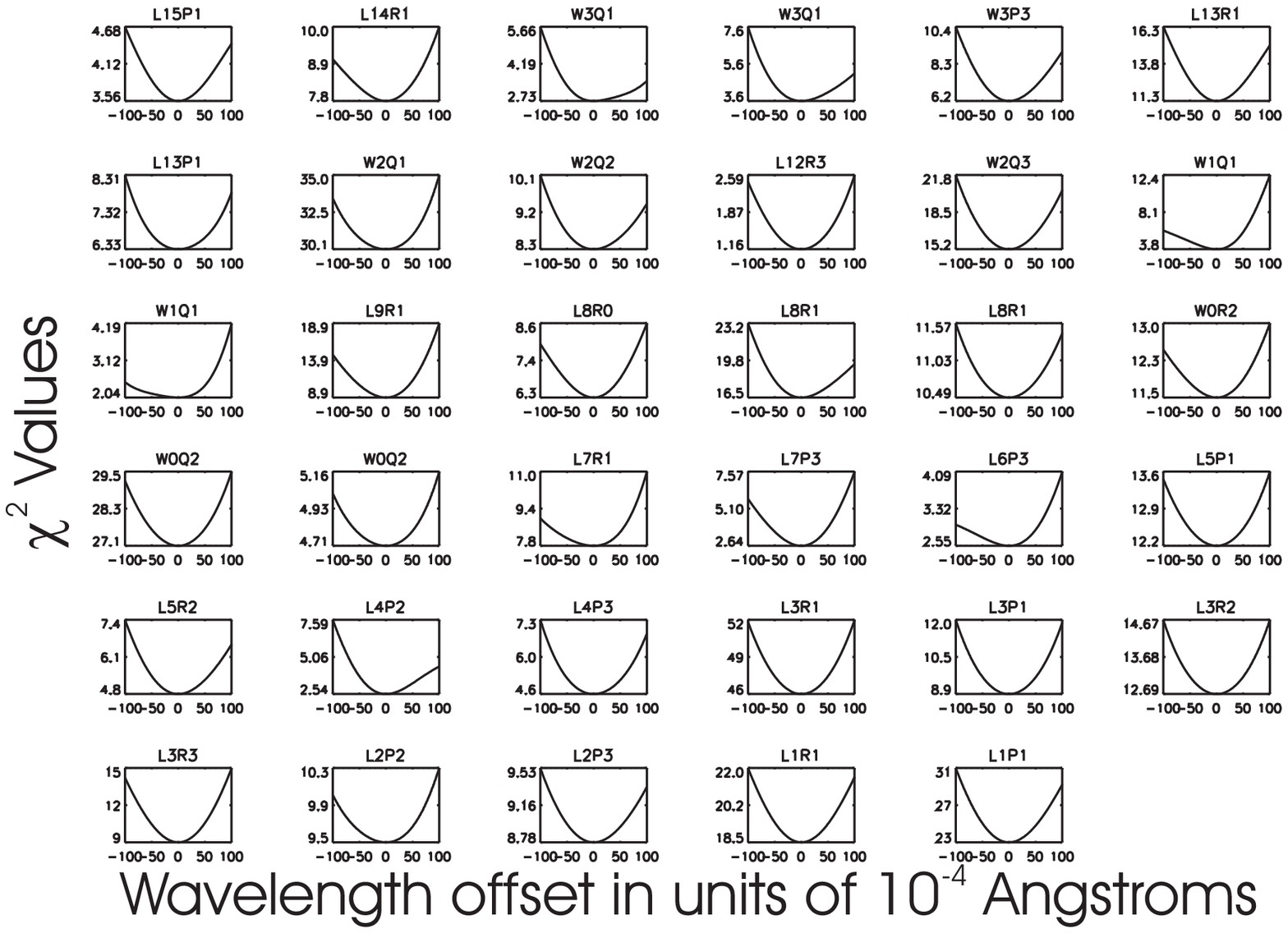}
\caption{The $\chi^2$ values for a sweep of wavelengths in $10^{-4}$ \AA\
increments around the line wavelength at the minimum $\chi^2$ value
for Q0347-383. Note that the actual analysis used wavelength
increments of $10^{-5}$ \AA\ rather than $10^{-4}$.
Transitions are labeled in the same manner as in Tables~\ref{tab-d47}
and \ref{tab-d45}. \label{fig-c4}}
\end{figure}

\begin{figure}
\plotone{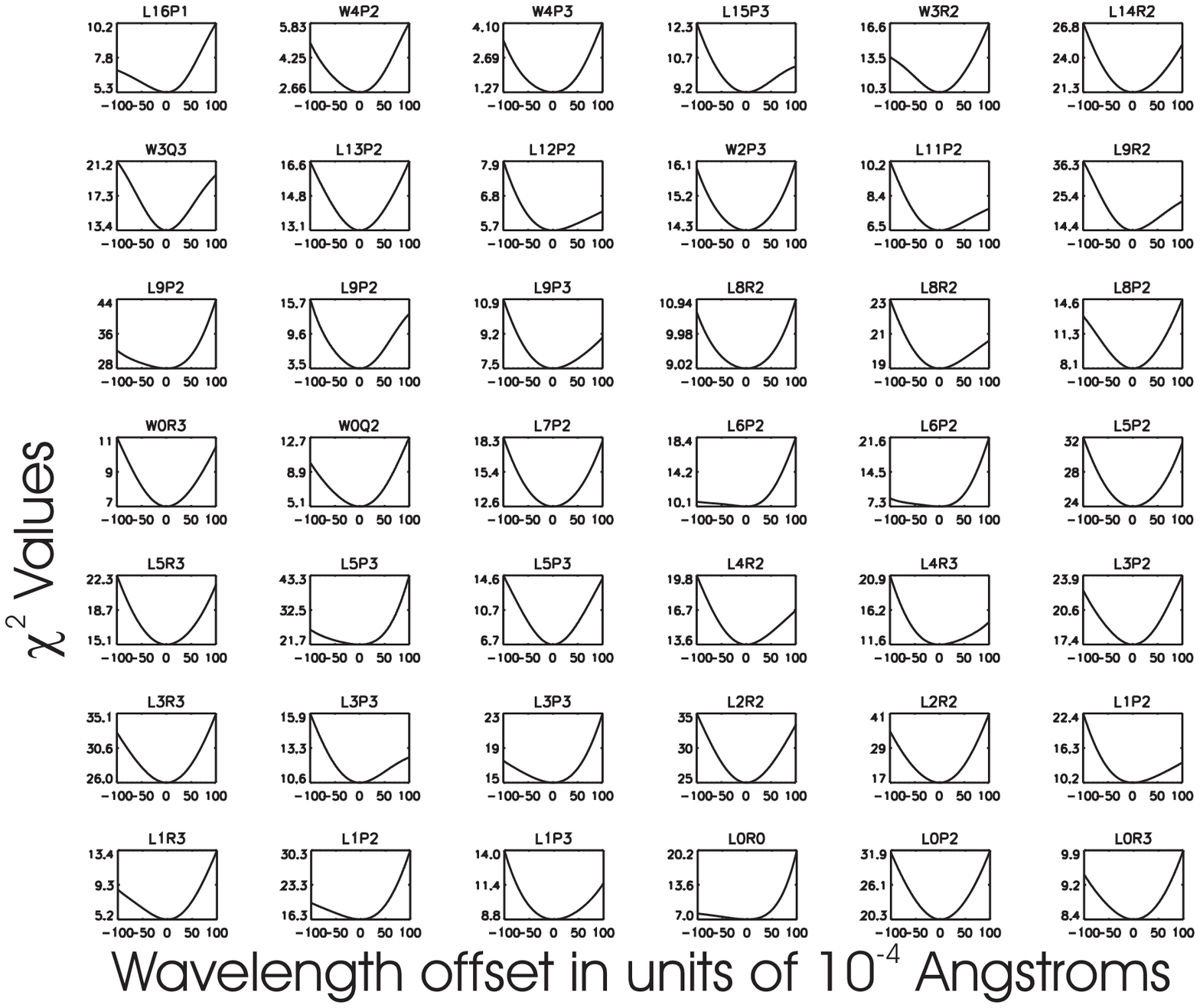}
\caption{The $\chi^2$ values for a sweep of wavelengths in $10^{-4}$ \AA\ 
increments around the line wavelength at the minimum $\chi^2$ value
for Q0405-443.  Note that the actual analysis used wavelength
increments of $10^{-5}$ \AA\ rather than $10^{-4}$.
Transitions are labeled in the same manner as in Tables~\ref{tab-d47}
and \ref{tab-d45}. \label{fig-c3}}
\end{figure}

Mosaic plots of the $\chi^2$ values calculated for each line are shown in
Figures~\ref{fig-c4} and \ref{fig-c3}.  The values are shown for 100 wavelengths
on either side of the best wavelength.  The wavelength values are spaced by
$10^{-4}$ \AA\ rather than the $10^{-5}$ \AA\ spacing in the actual analysis.
Note that the $\chi^2$ values are smoothly varying in a semi-parabolic shape
with a definite minimum.  \citet{mwf08} point out that this is a necessary
criterion for valid $\chi^2$ values. Although the values are generally symmetric 
about the minimum there is significant asymmetry in some of the plots.  This is 
expected from line profiles that are also not symmetric. Only the wavelength, 
which is the primary parameter of interest, is varied in the plots, as opposed
to both the wavelength and the column density in the fitting procedure, so that 
the minimum $\chi^2$ value is for one degree of freedom.  The $\chi^2$ values 
range from values almost as low as 1.1 to values as high as 45.6, with 
Q0347-383 having in general lower $\chi^2$ values than Q0405-443 which
has a lower signal to noise spectrum.  There does not appear to be any
correlation between the individual $\chi^2$ values and the deviation of the
reduced redshift from the average reduced redshift for each object. The
lines with the highest deviation from the average reduced redshift are not
the ones with the highest $\chi^2$ values.

\begin{figure}
\plotone{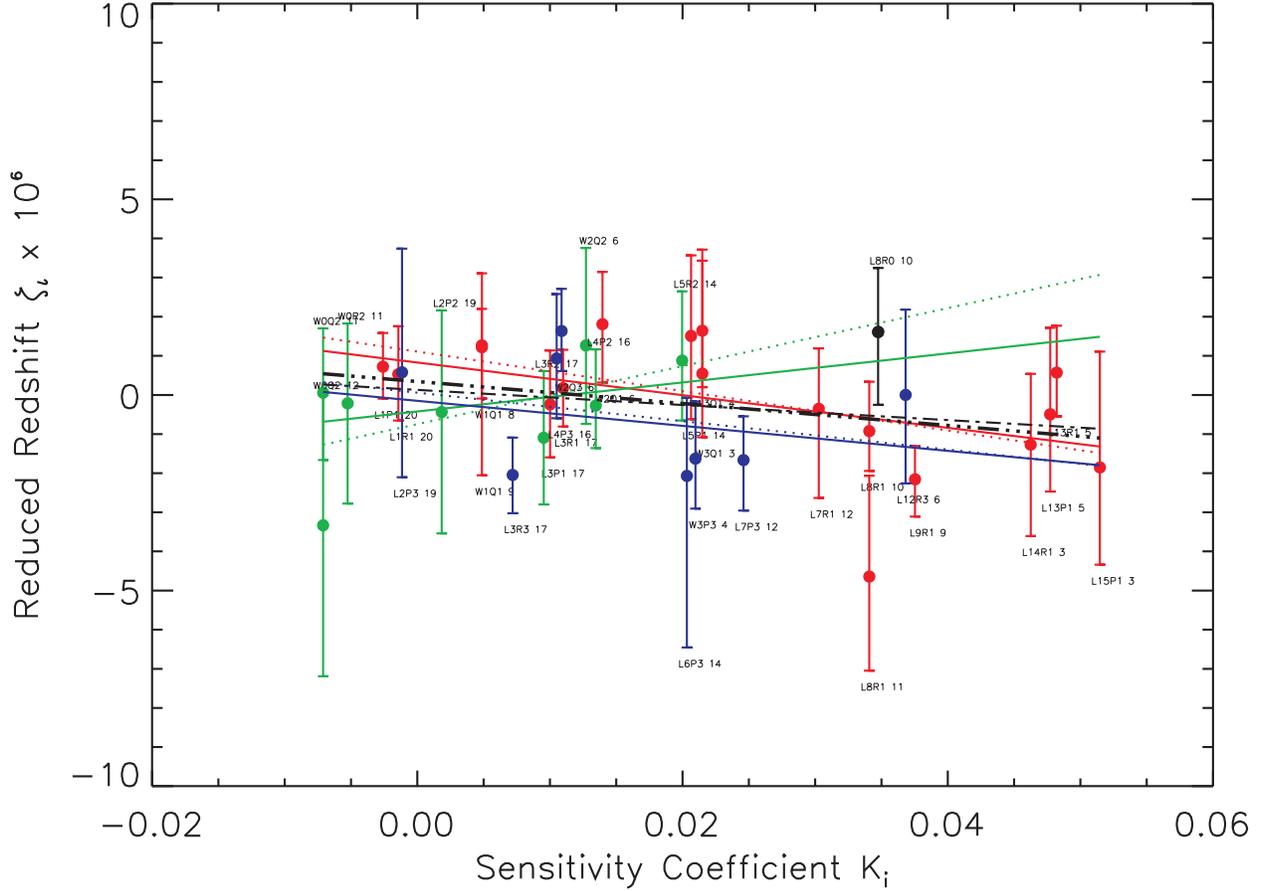}
\caption{The reduced redshift versus sensitivity factor plot for Q0347-383.
In the electronic version the symbols are color coded according to the rotational
level of the lower electronic state. J=0(black), J= 1(red), J=2(green), 
J=3(blue).  The solid line is the weighted fit and the dotted line is the 
unweighted fit to the individual J levels.  The thick dash 3 dot line is the
weighted fit and the thick dash dot line the unweighted fit to all J levels
combined.  The transitions are labeled with the last number being the order.
The orders are the observed order with the true order being 126 minus the 
printed number. \label{fig-a}}
\end{figure}

\begin{figure}
\plotone{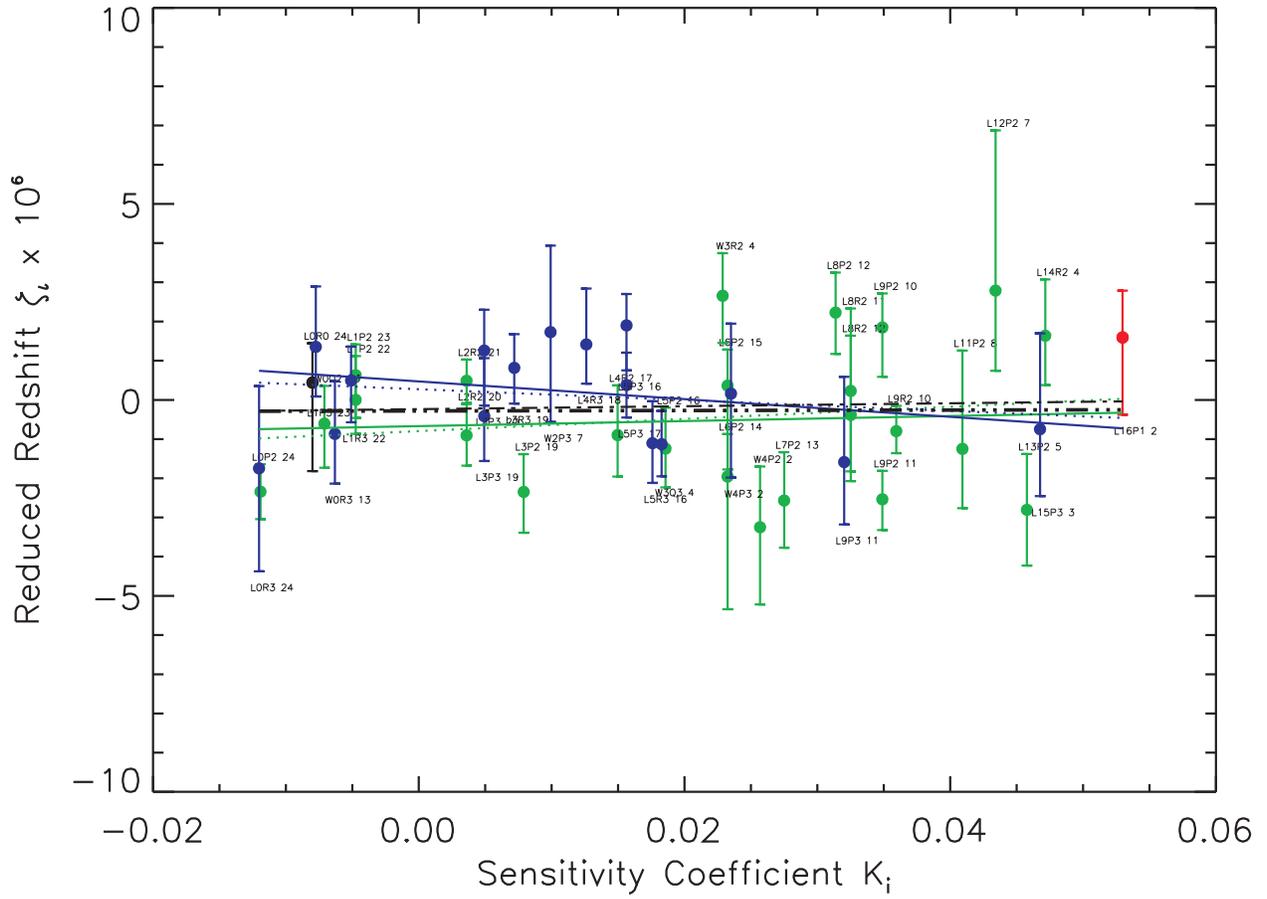}
\caption{Same as Figure~\ref{fig-a} except for Q0405-443.  In this plot
the true order number is 141 minus the printed number. \label{fig-b}}
\end{figure}

The error bars for each line in Figures \ref{fig-a} and \ref{fig-b} are
calculated by running the line fit calculations on either side of the 
minimum $\chi^2$ wavelength until the $\chi^2$ value increases by unity 
over the minimum $\chi^2$ value.  This produces in some cases a significant
difference between the positive and negative error bars.  The positive and
negative errors along with the $\chi^2$ values for the individual line fits
are listed in Tables~\ref{tab-d47} and \ref{tab-d45}.

\subsection{Results} \label{ss-res}

Tables~\ref{tab-d47} and \ref{tab-d45} give the results of the line fitting
for Q0347-383 and Q0405-443 respectively. Note that some transitions are 
repeated since they appear in two different orders.  In our analysis we 
treat these as independent measures.  Figures~\ref{fig-a} and \ref{fig-b}
are the plots of reduced redshift $\zeta$ versus sensitivity factor K$_i$ 
for Q0347-383 and Q0405-443 in manner similar to that first used by \citet{var93}.  
\citet{rei06} and \citet{uba07} both display their results with similar
plots.  The reduced redshift $\zeta$ is defined by 

\begin{equation}\label{eq-rz}
\zeta_i = \frac{z_i - z_Q}{1 + z_Q} = \frac{\Delta \mu}{\mu} K_i
\end{equation}

\noindent where $z_Q$ is the true redshift of the system taken as the median 
redshift of all of the lines, $z_i$ is the redshift of individual lines and
$K_i$ is defined in Equation~\ref{eq-k}.  The median redshifts of the 
H$_2$ absorptions
in Q0347-383 and Q0405-443 are 3.0248996 and 2.5947366 respectively, relative
to the earth - sun barycenter. The slope in this plot is the value of 
$\Delta \mu / \mu$.

\begin{figure}
\plotone{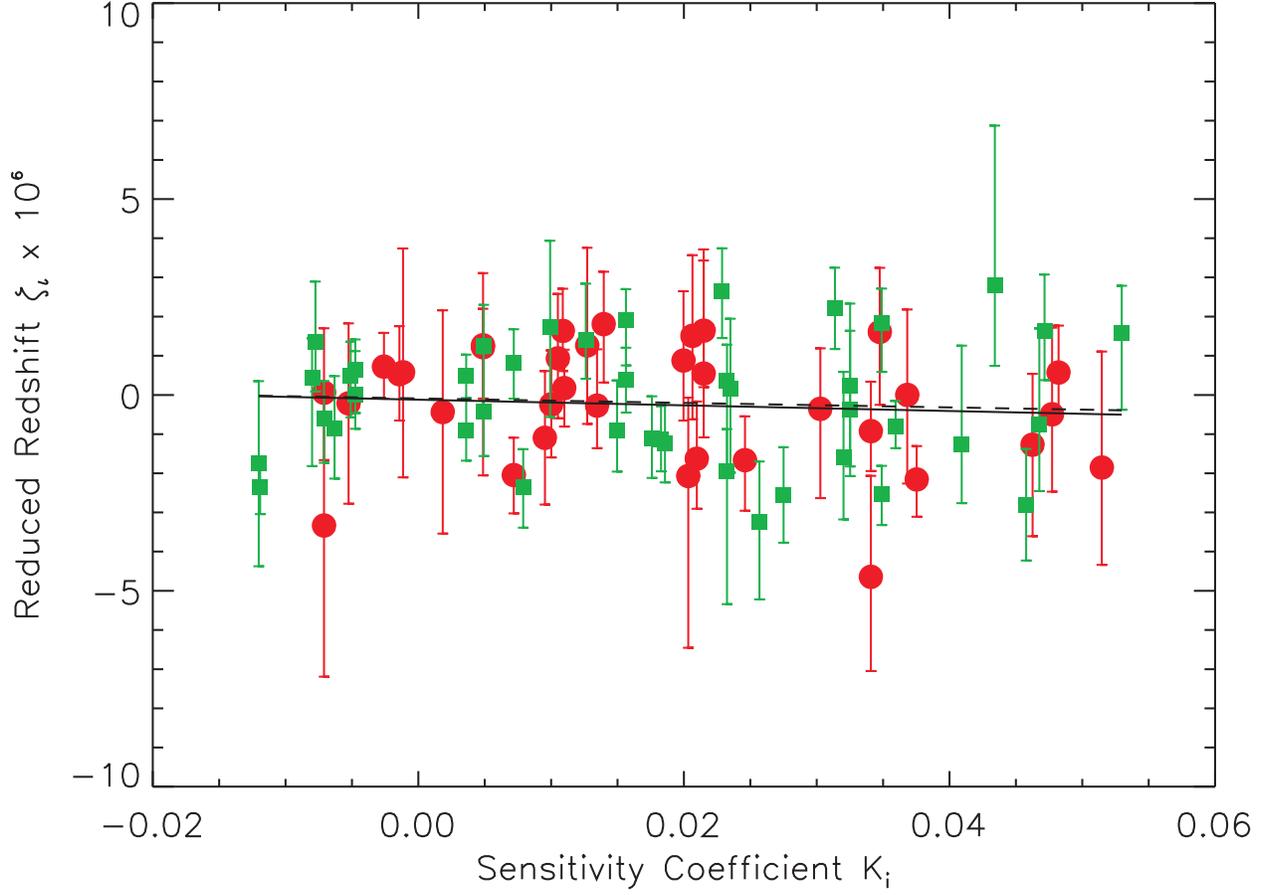}
\caption{The combined data plot of the reduced redshift $\zeta$ versus the
sensitivity parameter K.  The red dots are for Q0347-383 and the green squares
are for Q0405-443. The error bars are $1\sigma$. The dashed line is the 
unweighted fit to the data, $\Delta \mu / \mu = -6 \times 10^{-6} \pm 
10. \times 10^{-6}$ and the solid line the weighted fit to the data, 
$\Delta \mu / \mu = -7 \times 10^{-6} \pm 8 \times 10^{-6}$  \label{fig-cd}}
\end{figure}

The thick black dash dot and dash triple dot lines in the figures are the
weighted and unweighted linear least squares fits to the combined data for 
all rotational levels for each object where the weights are determined by
the standard deviations of the individual data points. The colored 
light solid and dotted lines are the fits to only lines with the same 
rotational level ground states given by the color codes in the caption.
The weighted and unweighted 
fits to Q0347-383 are $\Delta \mu / \mu = (-28 \pm 16) \times 10^{-6}$ and 
$\Delta \mu / \mu = (-19 \pm 15) \times 10^{-6}$.  The weighted and 
unweighted fits for Q0405-443 are $\Delta \mu / \mu = (0.55 \pm 10) 
\times 10^{-6}$ and $\Delta \mu / \mu = (3.7 \pm 14) \times 10^{-6}$.  
For the combined data set shown in Figure~\ref{fig-cd} the weighted fit yields
$\Delta \mu / \mu = (-7.0 \pm 8) \times 10^{-6}$ and the unweighted fit 
gives $\Delta \mu / \mu = (-6 \pm 10) \times 10^{-6}$. Both of these results 
are consistent with no variation in $\mu$ at the $68\%$ confidence level.  
Our result is consistent with the findings of \citet{wen08} giving
$-7.0 \times 10^{-6} \leq \Delta\mu/\mu \leq 49 \times 10^{-6}$ and 
\citet{kin09} which give $\Delta\mu/\mu = (2.6 \pm 3.0) \times 10^{-6}$. 
They are inconsistent at a roughly $3\sigma$ level with that of \citet{rei06} 
and \citet{uba07} who found $\Delta\mu/\mu = (24.5 \pm 5.9) \times 10^{-6}$.

The stated errors between the three measurement vary widely.  It should
be noted that the error quoted by \citet{wen08} is a $2\sigma$ error so 
the value $\pm 25 \times 10^{-6}$ should be used in comparison to the
other two measurements.  \citet{kin09} perform an analysis where they use
the same lines as \citet{uba07} and their error grows to about $\pm 9 \times
10^{-6}$ for the individual objects.  This error is similar but slightly
smaller than our errors.  \citet{kin09}  also
include a third object Q0528-250 which has a significantly smaller quoted
error than the two objects common to all of the studies.  In the weighted
mean of errors Q0528-250 has a dominant effect on the quoted error.
In the following we discuss our error analysis.

\subsection{Error Analysis} \label{ss-ea}

The least squares linear fit to the weighted combined data of 77 lines gives 
a $\chi^2$ value of 104.9.  Assuming N-2 degrees of freedom this is a 
$\chi^2$ value of 1.4 per degree of freedom.  In the combined data there are 
10 pairs of lines of the same transition but observed in a different
orders. Q0347-383 has 4 pairs and Q0405-443 has 6 pairs.  We have treated these
as independent measurements since they have independent photon noise and read
noise statistics.  Systematic effects such as unknown blended lines and 
continuum shape may introduce systematics into the measurements.  Individual
inspection of the line pairs indicates that the dispersion in reduced redshift
between the two measurements is consistent with the dispersion between 
independent lines and should not bias the results.

\subsubsection{Bootstrap Analysis}

\begin{figure}
\plotone{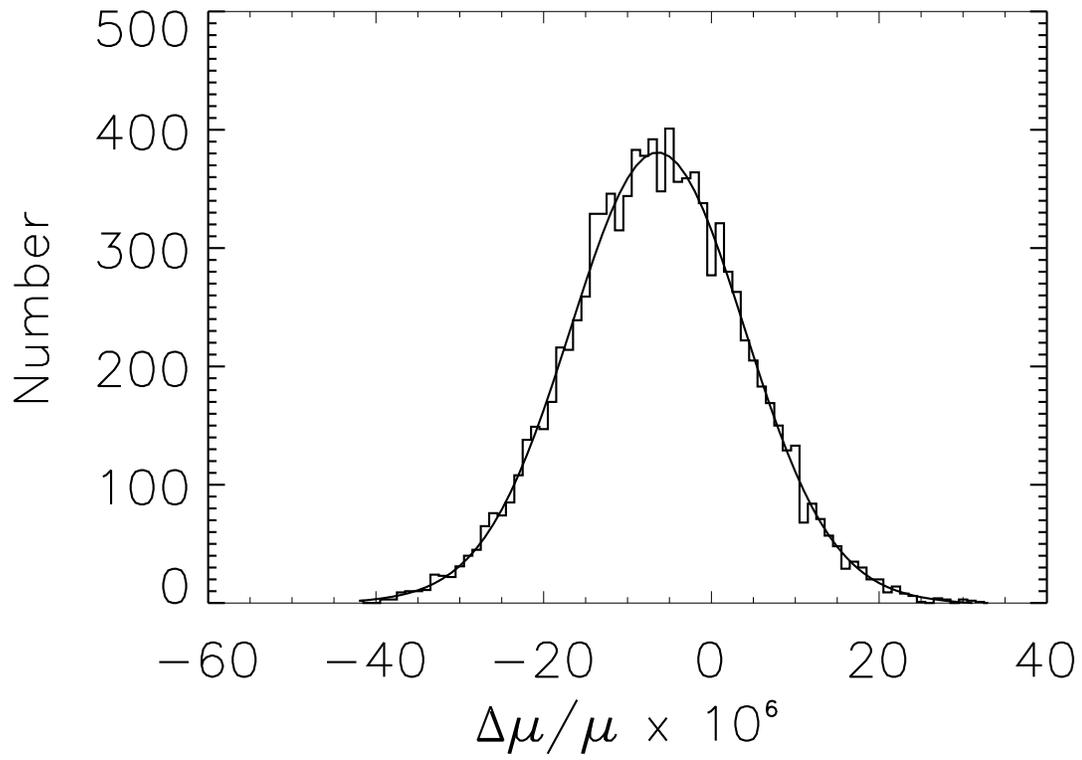}
\caption{The histogram is the output of a 10,000 sample boot strap analysis of
the original data.  The histogram bins are unity in the units of the
abscissa and the ordinate is the number of samples in the bin. The smooth
curve is a Gaussian fit to the histogram. \label{fig-bs}}
\end{figure}

As an alternative check on the statistical significance of the null result we
performed a bootstrap analysis on the combined data set.  We produced 10,000
new data sets by drawing the same number of lines but randomly selected allowing
duplication from the original data set.  Linear least square fits were performed
on the data sets and the result plotted as a histogram shown in Figure~\ref{fig-bs}.
The smooth curve in the figure is a Gaussian fit to the histogram.  The peak
of the Gaussian is at a $\Delta\mu/\mu$ value of $-6.4 \times 10^{-6}$ and the
half width at half maximum is $12 \times 10^{-6}$, both of which are consistent
with the results of the $\chi^2$ analysis.  The histogram values conform to the
Gaussian fit quite well indicating the appropriateness of the assumption of
Gaussian distributed errors.

\subsubsection{Ground State Rotational Levels}

Since transitions from 4 different rotational levels (J=0,1,2,3) are used it
is possible that the transitions could arise from physically offset regions
of the molecular cloud that could also have velocity offsets.  Our null result
is less prone to this type of error, however, velocity offsets could be in
a direction to reduce a $\Delta\mu/\mu$ signal.  To check for this the solutions
for lines with the same rotational ground level and the same object are plotted
in Figures~\ref{fig-a} and \ref{fig-b}.  In Q0347-383 there is only one line
with a ground rotational level of 0 so no solution is plotted for it.  The 
J=1 and 3 solutions have roughly the same slope while the J=2 solution has a
different slope with the opposite sign. This would appear difficult to achieve
with physical velocity offsets which would be presumed to have a smooth gradient
of velocity with temperature.  In Q0405-443 there are no J=0 lines and only one
J=1 line so no solutions exist for those systems.  The J=2 and J=3 slopes are
of opposite sign but within the error bars of each other.  From this analysis
we conclude that it is unlikely that velocity gradients with excitation level 
are masking a change in $\mu$.

\subsubsection{Combination of the Data Sets}

To improve the statistics of the sample we have combined the lines from the 
two systems.  The higher redshift system associated with Q0347-383 shows
a shift in $\mu$ at the $1\sigma$ level in the unweighted fit and at 
a $1.75\sigma$ level in the weighted fit, both indicating a decrease in the 
value of $\mu$.  The system associated with Q0405-443 at a lower redshift
shows an increase in $\mu$ in both the unweighted and weighted fit but at
levels significantly less that $1\sigma$.  The Q0405-443 system has 7 more
lines than the Q0347-383 system.  It could be argued that the combination
of the two data sets dilutes the signal of a real shift in the Q0347-383
system at the higher redshift and earlier time in the universe.  Although
we would not claim a real shift in $\mu$ with a $1.75\sigma$ result we can
not rule out that the combined data set is diluting the evidence for a
change in $\mu$.  The higher than expected value of the $\chi^2$ per degree
of freedom could be due to the difference in fitted slopes of the two systems
taken separately.

\subsubsection{Systematics}

The method of sensitivity coefficient fitting is subject to systematic 
errors in the wavelength scale.  In general the sensitivity coefficients
increase with increasing vibrational energies in the upper level of the
transitions.  This means that for a given electronic transition system
the sensitivity coefficient increases with decreasing transition 
wavelength.  That means that any systematic error that produces an
erroneous gradient in the wavelength calibration will mimic a change
in $\mu$.  This is mitigated to some degree by the mixture of Lyman
and Werner bands.  The higher electronic energy of the upper level of
the Werner system places low values of the sensitivity coefficient at
the same wavelengths as high sensitivity coefficient Lyman transitions.
At wavelengths longer than the longest wavelengths of the Werner system,
however, there is no mitigating effect.  It may be this effect coupled
with the systematic errors in the older UVES pipeline reductions found
by \citet{mur08} that produced the positive detection of a change in
$\mu$ by \citet{rei06}. The analysis in \citet{thm09}, Figures 7 and 8, 
indicates that 
the wavelength calibration used in this analysis is not subject to
systematic errors of the magnitude cited in \citet{mur08}.  In addition
our wavelength calibration is on an order by order basis which resets
the solution for each order, making it more difficult to have systematic
effects over the whole wavelength solution.

\subsubsection{Comparison of Lyman and Werner Lines}

\begin{figure}
\plottwo{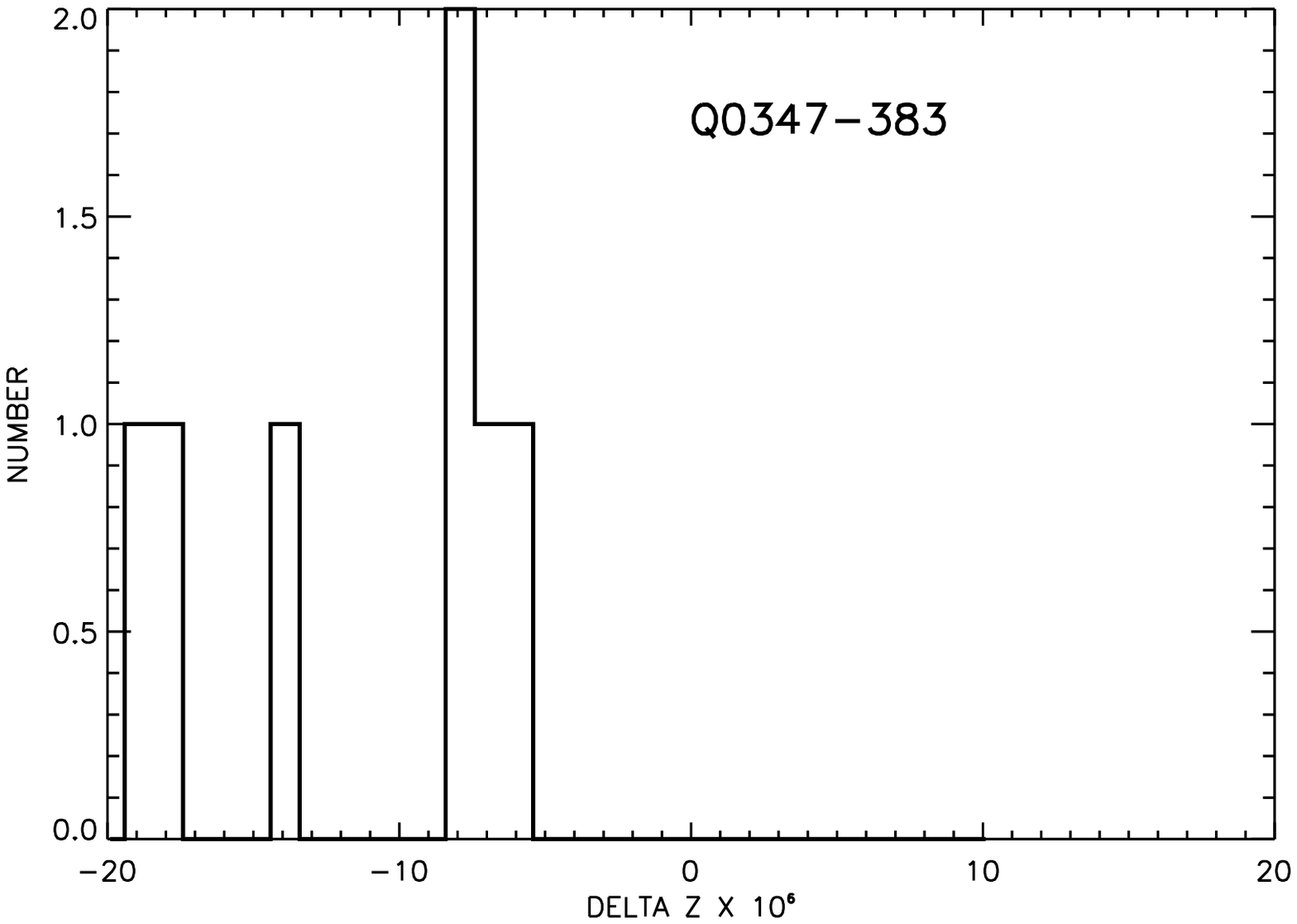}{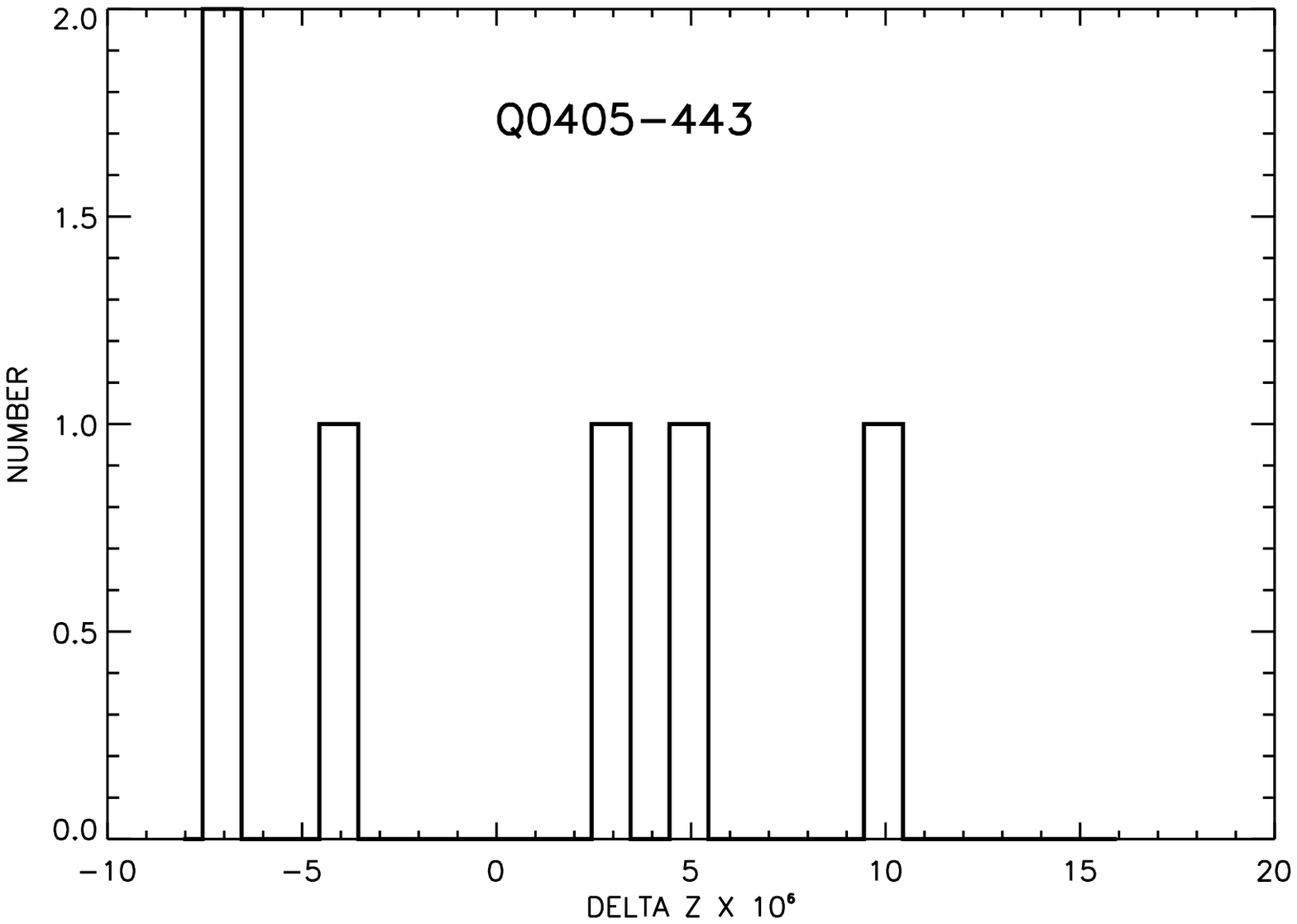}
\caption{Histograms of the delta redshift between each of the Werner band lines
and their adjacent Lyman band lines.  A positive value means that the Lyman
line had a higher redshift than the Werner line.  The left histogram is for
Q0347-383 and the right is for Q0405-443. Note that this is the delta redshift not
the delta reduced redshift defined in Equation~\ref{eq-rz}. \label{fig-wl}}
\end{figure}

The Werner lines with a higher upper electronic level provide lines with
low upper state vibrational levels at wavelengths that are close to Lyman
lines with high upper state vibrational levels.  Since the sensitivity factors
are roughly proportional to upper state vibrational level this mixes lines 
with low sensitivity factors with those with high factors.  Under the
assumption that any possible systematic wavelength errors are minimized
for lines that lie close together we have looked at the redshift differences
between all of the Werner lines and the Lyman lines that are adjacent to
them in the same order.  There are a limited number of line pairs that satisfy
this criterion, 8 for Q0347-383 and 7 for Q0405-443.  Histograms of the
distribution are given in Figure~\ref{fig-wl}.  The distribution of delta
redshift values (z(Lyman) - z(Werner)) for Q0405-443 are roughly evenly 
distributed around zero but the delta redshift values for Q0347-383 are
all negative.  This is consistent with the negative slope of the fit in
Figure~\ref{fig-a}.

\subsection{The Marginal Possibility of a Shift in Q0347-383}

The analysis results for Q0347-383 show a negative shift in the value
of $\mu$ at the 1.75 $\sigma$ level which has a statistical probability of
being a true shift at the 91$\%$ percent level if the errors are Gaussian
distributed.  This is certainly not a
level which justifies declaring a change in a fundamental constant
but raises the marginal possibility that there might be a change. In
addition the comparison between the Werner and Lyman lines in 
Figure~\ref{fig-wl} shows a negative delta between the Werner and
Lyman lines for a seven cases which has a probability of 
$2^{-7} = 0.008$ chance of happening randomly. If a monotonically
rolling scalar field is invoked for the change, the higher redshift
of Q0347-383 could be why a change is seen in Q0347-383 and not
Q0405-443.  We consider this evidence as suggestive but in no way
conclusive. It does point out the need for observations of systems
at higher redshift.

\section{Conclusions and Implications} \label{s-con}

Our basic conclusion, based on the combination of data from Q0347-383
and Q0405-443, is that there has been no change in the value of $\mu$
to 1 part in $10^5$ over a time span of 11.5 gigayears.  This is approximately
$80\%$ of the age of the universe.  The accuracy of the limit on
$\Delta\mu/\mu$ is set by both the spectral resolution and the signal 
to noise ratio of the flux.  This conclusion is consistent with the results
of \citet{kin09} but inconsistent with the results of \citet{rei06}.
Starting with the same raw data, the primary difference in this 
analysis is the use of improved wavelength calibration techniques that 
eliminated the systematic variations in the calibration used in the UVES 
pipeline at the time of the \citet{rei06} analysis. The line selection is
also most likely different from \citet{rei06} but without a list of those
lines it is difficult to assess the influence of the lines chosen. There 
is a marginal possibility of the detection of a 
change in $\mu$ based on the Q0347-383 data alone.  We, however,
feel that this result while suggestive is certainly not conclusive.

What implications does a limit on $\Delta\mu/\mu$ of $10^{-5}$ have on
theories of dark energy that invoke a rolling scalar field potential
as the source of the dark energy? \citet{cho07} have despaired about 
distinguishing between a universe with a cosmological constant relative
to a universe with a quintessence rolling scalar field, however, the 
former predicts no change in $\mu$ while the latter predicts a change
even though the magnitude or even the sign of the change is not presently
calculable.  Detection of a change in $\mu$ or its companion the fine
structure constant $\alpha$ would be strong evidence for quintessence 
as opposed to a cosmological constant.

Quintessence is usually expressed in terms of a potential $V(\phi)$ that
is a function of the rolling scalar $\phi$.  The change in $\mu$ is 
then expressed as

\begin{equation} \label{eq-dm}
\frac{\Delta \mu}{\mu} = \zeta_{\mu}\kappa(\phi - \phi_0)
\end{equation}

\noindent where $\kappa$ is $\sqrt{8 \pi} / m_{Pl}$, $m_{Pl}$ is
the Planck mass and $\zeta_{\mu}$ is
a parameter of unknown value (\citet{ave06} and references therein). 
Determination of the value of $\mu$ at high redshift is therefore a 
direct way to distinguish between quintessence and a cosmological
constant. In Grand Unified Theories (GUTs) the rolling of $\mu$ is 
typically given by

\begin{equation} \label{eq-gut}
\frac{\dot{\mu}}{\mu} \sim \frac{\dot{\Lambda}_{QCD}}{\Lambda_{QCD}}-
\frac{\dot{\nu}}{\nu} \sim R\frac{\dot{\alpha}}{\alpha}
\end{equation}

\noindent where $\Lambda_{QCD}$ is the QCD scale, $\nu$ is the Higgs 
Vacuum Expectation Value (VEV), R is a model dependent value
(\citet{ave06} and references therein) and $\alpha$ is the fine structure
constant.  In many GUT models the value
of R is large and negative $\sim-50$ \citet{ave06}.

Our current results limit the value of $\zeta_{\mu}\kappa(\phi-\phi_0)$
in Equation~\ref{eq-dm} to be on the order of $10^{-5}$ or less, but
does not tell us the individual values of $(\phi - \phi_0)$ or 
$\zeta_{\mu}$.  The results do, however, rule out Model A of
\citet{ave06} at about the 4$\sigma$ level where the potential is
given by

\begin{equation} \label{eq-a}
V(\phi) = V_0(\exp(10\kappa \phi)  + \exp(0.1\kappa \phi))
\end{equation}

\noindent which predicts a value of 
$\frac{\Delta \mu}{\mu} = 3 \times 10^{-5}$ at a redshift of 3.
This means that even at the current level of accuracy significant
bounds on the quintessence models are being established.  In all
fairness to the model it must be pointed out that it was designed
to achieve that result to match the findings of \citet{rei06}.

If the claim of a detected change in the fine structure constant
$\alpha$ ($\frac{\Delta \alpha}{\alpha} = 0.57 \times 10^{-5}$ 
\citet{mwf03}) is accepted then this implies a value of R of
$\le 2$ which is significantly different in sign and magnitude
that the typical GUT value quoted above.  This would mean that
either the roll of both the QCD scale and the Higgs VEV is small
or that they are equal to each other by less than a factor of 2.
Of course if the claim for a change in $\alpha$ is not accepted 
the current limitation on $\frac{\Delta \mu}{\mu}$ places no limit
on R.

\acknowledgments

RIT would like to acknowledge interesting and useful conversations
with Wim Ubachs, Dimitrios Psaltis, Feryal Ozel and Michael Murphy on theory
and technique. C.M. wishes to acknowledge very useful discussions 
with Paolo Molaro.  The work of C.M. is funded by a Ciencia2007
research grant.

\clearpage

\begin{deluxetable}{cccc}
\tabletypesize{\scriptsize}
\tablecaption{Recent Astronomical $\mu$ Measurements \label{tab-comp}}
\tablewidth{0pt}
\tablehead{
\colhead{Object} & \colhead{Reference} & \colhead{Redshift} & \colhead{$\Delta\mu/\mu$}
}
\startdata
PKS 0528-250 & \citet{fol88} & 2.811 & $| | \le 2 \times 10^{-4}$ \\
PKS 0528-250 & \citet{cow95} & 2.811 & $| | \le 7 \times 10^{-4}$ \\
PKS 0528-250 & \citet{pot98} & 2.811 & $| | \le 2 \times 10^{-4}$ \\
Q0347-383 + Q1232+082 & \citet{iva02} & 3.0249 & $(5.7 \pm 3.8) \times 10^{-5}$ \\
Q0347-383 & \citet{lev02} & 3.0249 & $-1.5 \times 10^{-5} \le 5.7 \times 10^{-5}$ \\
Q0347-383 & \citet{iva03} & 3.0249 & $| | \le 8 \times 10^{-5}$ \\
Q0347-383 & \citet{wen08} & 3.0249 & $-0.7 \times 10^{-5} \le 4.9 \times 10^{-5}$ \\
Q0347-383 + Q0405-443 & \citet{uba04} & 3.0249, 2.5974 & $(-0.5 \pm 3.8) \times 10^{-5}$ \\
Q0347-383 + Q0405-443 & \citet{iva05} & 3.0249, 2.5974 & $(1.64 \pm 0.74) \times 10^{-5}$ \\
Q0347-383 + Q0405-443 & \citet{rei06} & 3.0249, 2.5974 & $(2.4 \pm 0.6) \times 10^{-5}$ \\
Q0347-383 + Q0405-443 & \citet{rei06} & 3.0249, 2.5974 & $(2.45 \pm 0.59) \times 10^{-5}$ \\
Q0347-383 + Q0405-443 & this work & 3.0249, 2.5974 & $(-7 \pm 8) \times 10^{-6}$ \\
Q0347-383 + Q0405-443 + PKS 0528-250 & \citet{kin09} & 3.0249, 2.5974, 2.811 & $(2.6 \pm 3.0) \times 10^{-6}$ \\
B0218+357 & \citet{fla07} & 0.6847 & $(0.6 \pm 1.9) \times 10^{-6}$ \\
B0218+357 & \citet{mur08b} & 0.6847 & $| | \le 0.18 \times 10^{-6}$ \\
Milky Way & \citet{lev08} & 0.0 & $(4 - 14) \times 10^{-8}$ \\
\enddata
\end{deluxetable}

\clearpage

\begin{deluxetable}{ccc}
\tabletypesize{\scriptsize}
\tablecaption{Observational Parameters for Q0347-383 \label{tab-obs7}}
\tablewidth{0pt}
\tablehead{
\colhead{Archive File} & \colhead{Date} & \colhead{Seconds Exp.} 
}
\startdata
UVES\_2002\_01\_08T00:46:05\_351\_b.fits & 8 Jan. 2002 & 4500\\
UVES\_2002\_01\_08T02:03:41\_018\_b.fits & 8 Jan. 2002 & 4500\\
UVES\_2002\_01\_08T03:21:18\_348\_b.fits & 8 Jan. 2002 & 4500\\
UVES\_2002\_01\_09T00:43:43\_109\_b.fits & 9 Jan. 2002 & 4500\\
UVES\_2002\_01\_09T02:02:11\_833\_b.fits & 9 Jan. 2002 & 4500\\
UVES\_2002\_01\_09T03:19:58\_841\_b.fits & 9 Jan. 2002 & 4500\\
UVES\_2002\_01\_10T00:48:56\_171\_b.fits & 10 Jan. 2002 & 4500\\
UVES\_2002\_01\_10T02:06:28\_725\_b.fits & 10 Jan. 2002 & 4500\\
UVES\_2002\_01\_10T03:24:33\_981\_b.fits & 10 Jan. 2002 & 4500\\
\enddata

\end{deluxetable}

\clearpage

\begin{deluxetable}{ccc}
\tabletypesize{\scriptsize}
\tablecaption{Observational Parameters for Q0405-443 \label{tab-obs5}}
\tablewidth{0pt}
\tablehead{
\colhead{Archive File} & \colhead{Date} & \colhead{Exposure Time} 
}
\startdata
UVES\_2003\_01\_04T00:43:06\_274\_b.fits & 4 Jan. 2003 & 4500\\
UVES\_2003\_01\_04T02:09:06\_464\_b.fits & 4 Jan. 2003 & 4500\\
UVES\_2003\_01\_04T03:34:08\_623\_b.fits & 4 Jan. 2003 & 4500\\
UVES\_2003\_01\_05T00:48:35\_827\_b.fits & 5 Jan. 2003 & 4500\\
UVES\_2003\_01\_05T02:16:14\_922\_b.fits & 5 Jan. 2003 & 4500\\
UVES\_2003\_01\_05T03:46:36\_522\_b.fits & 5 Jan. 2003 & 4500\\
UVES\_2003\_01\_06T00:45:18\_207\_b.fits & 6 Jan. 2003 & 4500\\
UVES\_2003\_01\_06T02:15:26\_790\_b.fits & 6 Jan. 2003 & 4500\\
UVES\_2003\_01\_06T03:46:49\_242\_b.fits & 6 Jan. 2003 & 4500\\
\enddata
\end{deluxetable}

\clearpage
\begin{deluxetable}{lcccccccc}
\tabletypesize{\scriptsize}
\tablecaption{Q0347-383 Line List \label{tab-d47}}
\tablewidth{0pt}
\tablehead{
\colhead{Trans.}\tablenotemark{a} & \colhead{order} & \colhead{K factor} & 
\colhead{Obs. wavelength}\tablenotemark{b} &\colhead{pos. error} & 
\colhead{neg. error} & \colhead{$\chi^2$} & \colhead{Rest wavelength}
\tablenotemark{b} & \colhead{redshift}\tablenotemark{c}}
\startdata
  L15P1 &123 & 0.05147000 & 3782.21819 &  0.0112 & -0.0094 & 3.56 &  939.70672 & 3.02489213\\
  L14R1 &123 & 0.04625000 & 3811.49618 &  0.0069 & -0.0089 & 7.84 &  946.98040 & 3.02489448\\
   W3Q1 &123 & 0.02149000 & 3813.28002 &  0.0110 & -0.0062 & 2.73 &  947.42188 & 3.02490179\\
   W3Q1 &122 & 0.02149000 & 3813.28420 &  0.0079 & -0.0055 & 3.59 &  947.42188 & 3.02490620\\
   W3P3 &122 & 0.02097000 & 3830.37745 &  0.0056 & -0.0049 & 6.23 &  951.67186 & 3.02489304\\
  L13R1 &121 & 0.04821000 & 3844.04623 &  0.0046 & -0.0043 &11.31 &  955.06582 & 3.02490189\\
  L13P1 &121 & 0.04772000 & 3846.62792 &  0.0085 & -0.0076 & 6.33 &  955.70827 & 3.02489760\\
   W2Q1 &120 & 0.01396000 & 3888.44675 &  0.0052 & -0.0058 &30.09 &  966.09608 & 3.02490687\\
   W2Q2 &120 & 0.01272000 & 3893.21423 &  0.0097 & -0.0078 & 8.29 &  967.28110 & 3.02490468\\
  L12R3 &120 & 0.03682000 & 3894.80256 &  0.0085 & -0.0088 & 1.16 &  967.67695 & 3.02489959\\
   W2Q3 &120 & 0.01088000 & 3900.33218 &  0.0042 & -0.0040 &15.21 &  969.04922 & 3.02490617\\
   W1Q1 &118 & 0.00487000 & 3971.76790 &  0.0037 & -0.0054 & 3.82 &  986.79800 & 3.02490469\\
   W1Q1 &117 & 0.00487000 & 3971.76771 &  0.0075 & -0.0130 & 2.04 &  986.79800 & 3.02490450\\
   L9R1 &117 & 0.03753000 & 3992.75767 &  0.0034 & -0.0038 & 8.86 &  992.01637 & 3.02489091\\
   L8R0 &116 & 0.03475000 & 4032.24698 &  0.0066 & -0.0075 & 6.32 & 1001.82387 & 3.02490607\\
   L8R1 &116 & 0.03408000 & 4034.76532 &  0.0051 & -0.0041 &16.47 & 1002.45210 & 3.02489587\\
   L8R1 &115 & 0.03408000 & 4034.75032 &  0.0104 & -0.0097 &10.49 & 1002.45210 & 3.02488091\\
   W0R2 &115 &-0.00525000 & 4061.22312 &  0.0083 & -0.0104 &11.53 & 1009.02492 & 3.02489873\\
   W0Q2 &115 &-0.00710000 & 4068.92599 &  0.0067 & -0.0070 &27.08 & 1010.93845 & 3.02489982\\
   W0Q2 &114 &-0.00710000 & 4068.91220 &  0.0138 & -0.0157 & 4.71 & 1010.93845 & 3.02488618\\
   L7R1 &114 & 0.03027000 & 4078.98076 &  0.0063 & -0.0093 & 7.84 & 1013.43701 & 3.02489816\\
   L7P3 &114 & 0.02460000 & 4103.38732 &  0.0046 & -0.0053 & 2.64 & 1019.50224 & 3.02489289\\
   L6P3 &112 & 0.02033000 & 4150.43809 &  0.0083 & -0.0182 & 2.55 & 1031.19260 & 3.02489126\\
   L5P1 &112 & 0.02064000 & 4178.48451 &  0.0086 & -0.0089 &12.21 & 1038.15713 & 3.02490566\\
   L5R2 &112 & 0.01997000 & 4180.62771 &  0.0074 & -0.0064 & 4.82 & 1038.69027 & 3.02490313\\
   L4P2 &110 & 0.01346000 & 4239.36224 &  0.0061 & -0.0046 & 2.54 & 1053.28426 & 3.02489850\\
   L4P3 &110 & 0.01051000 & 4252.19544 &  0.0070 & -0.0065 & 4.61 & 1056.47144 & 3.02490335\\
   L3R1 &109 & 0.01099000 & 4280.32103 &  0.0042 & -0.0042 &45.60 & 1063.46014 & 3.02490030\\
   L3P1 &109 & 0.01001000 & 4284.92877 &  0.0059 & -0.0058 & 8.87 & 1064.60539 & 3.02489862\\
   L3R2 &109 & 0.00953000 & 4286.49249 &  0.0073 & -0.0073 &12.69 & 1064.99481 & 3.02489519\\
   L3R3 &109 & 0.00719000 & 4296.48519 &  0.0041 & -0.0042 & 9.06 & 1067.47855 & 3.02489136\\
   L2P2 &107 & 0.00184000 & 4351.98530 &  0.0113 & -0.0135 & 8.78 & 1081.26603 & 3.02489783\\
   L2P3 &107 &-0.00115000 & 4365.24899 &  0.0138 & -0.0117 &18.47 & 1084.56034 & 3.02490192\\
   L1R1 &106 &-0.00143000 & 4398.14064 &  0.0054 & -0.0052 &22.80 & 1092.73243 & 3.02490172\\
   L1P1 &106 &-0.00259000 & 4403.45255 &  0.0038 & -0.0036 &22.80 & 1094.05198 & 3.02490250\\
\enddata
\tablenotetext{a}{Transitions are labeled with L or W for Lyman or
Werner, then the vibrational quantum number of the upper state, next R,
Q or P transitions and finally the rotational quantum number of
the lower state}
\tablenotetext{b}{Vacuum wavelength}
\tablenotetext{c}{Barycentric redshift}
\end{deluxetable}

\clearpage
\begin{deluxetable}{lcccccccc}
\tabletypesize{\scriptsize}
\tablecaption{Q0405-443 Line List \label{tab-d45}}
\tablewidth{0pt}
\tablehead{
\colhead{Trans.}\tablenotemark{a} & \colhead{order} & \colhead{K factor} & 
\colhead{Obs. wavelength}\tablenotemark{b} &\colhead{pos. error} & 
\colhead{neg. error} & \colhead{$\chi^2$} & \colhead{Rest wavelength}
\tablenotemark{b} & \colhead{redshift}\tablenotemark{c}}
\startdata
  L16P1 &139 & 0.05297000 & 3351.25678 &  0.0040 & -0.0066 & 5.30 &  932.26621 & 2.59474230\\
   W4P2 &139 & 0.02569000 & 3352.45726 &  0.0052 & -0.0066 & 2.66 &  932.60468 & 2.59472489\\
   W4P3 &139 & 0.02350000 & 3360.32455 &  0.0060 & -0.0072 & 1.27 &  934.79006 & 2.59473715\\
  L15P3 &138 & 0.04676000 & 3394.61670 &  0.0083 & -0.0058 &10.34 &  944.33046 & 2.59473389\\
   W3R2 &137 & 0.02287000 & 3404.62607 &  0.0037 & -0.0041 &13.42 &  947.11169 & 2.59474612\\
  L14R2 &137 & 0.04715000 & 3409.50987 &  0.0049 & -0.0043 &13.11 &  948.47125 & 2.59474246\\
   W3Q3 &137 & 0.01828000 & 3416.42562 &  0.0029 & -0.0028 & 5.69 &  950.39773 & 2.59473251\\
  L13P2 &136 & 0.04577000 & 3442.49784 &  0.0049 & -0.0049 &14.30 &  957.65223 & 2.59472649\\
  L12P2 &134 & 0.04341000 & 3473.51556 &  0.0142 & -0.0071 & 6.53 &  966.27550 & 2.59474659\\
   W2P3 &134 & 0.00992000 & 3488.92549 &  0.0077 & -0.0080 &28.26 &  970.56332 & 2.59474279\\
  L11P2 &133 & 0.04092000 & 3506.10669 &  0.0088 & -0.0053 & 3.54 &  975.34576 & 2.59473208\\
   L9R2 &131 & 0.03594000 & 3571.54986 &  0.0023 & -0.0020 & 7.52 &  993.55061 & 2.59473370\\
   L9P2 &131 & 0.03489000 & 3576.31685 &  0.0031 & -0.0045 & 9.02 &  994.87408 & 2.59474322\\
   L9P2 &130 & 0.03489000 & 3576.30117 &  0.0026 & -0.0028 &18.84 &  994.87408 & 2.59472746\\
   L9P3 &130 & 0.03202000 & 3586.92017 &  0.0078 & -0.0057 & 8.06 &  997.82718 & 2.59473087\\
   L8R2 &130 & 0.03251000 & 3609.06404 &  0.0076 & -0.0083 & 6.95 & 1003.98545 & 2.59473739\\
   L8R2 &129 & 0.03251000 & 3609.06183 &  0.0073 & -0.0052 & 5.13 & 1003.98545 & 2.59473519\\
   L8P2 &129 & 0.03137000 & 3614.13175 &  0.0037 & -0.0038 &12.62 & 1005.39320 & 2.59474457\\
   W0R3 &128 &-0.00631000 & 3631.14901 &  0.0049 & -0.0046 & 7.27 & 1010.13025 & 2.59473346\\
   W0Q2 &128 &-0.00710000 & 3634.05522 &  0.0035 & -0.0041 &23.82 & 1010.93845 & 2.59473440\\
   L7P2 &128 & 0.02750000 & 3653.90105 &  0.0045 & -0.0044 &15.12 & 1016.46125 & 2.59472734\\
   L6P2 &127 & 0.02324000 & 3695.76334 &  0.0040 & -0.0125 &21.67 & 1028.10609 & 2.59472954\\
   L6P2 &126 & 0.02324000 & 3695.77191 &  0.0034 & -0.0079 & 6.73 & 1028.10609 & 2.59473788\\
   L5P2 &125 & 0.01857000 & 3739.84184 &  0.0039 & -0.0037 &11.57 & 1040.36733 & 2.59473210\\
   L5R3 &125 & 0.01759000 & 3742.68792 &  0.0040 & -0.0038 &17.41 & 1041.15892 & 2.59473260\\
   L5P3 &125 & 0.01564000 & 3751.12621 &  0.0030 & -0.0043 &26.05 & 1043.50319 & 2.59474340\\
   L5P3 &124 & 0.01564000 & 3751.12050 &  0.0031 & -0.0031 &10.63 & 1043.50319 & 2.59473793\\
   L4R2 &124 & 0.01497000 & 3779.85697 &  0.0048 & -0.0040 &15.03 & 1051.49857 & 2.59473334\\
   L4R3 &123 & 0.01261000 & 3788.77180 &  0.0054 & -0.0038 &25.20 & 1053.97610 & 2.59474166\\
   L3P2 &122 & 0.00790000 & 3835.21789 &  0.0037 & -0.0040 &17.12 & 1066.90068 & 2.59472813\\
   L3R3 &122 & 0.00719000 & 3837.30732 &  0.0033 & -0.0035 &10.22 & 1067.47855 & 2.59473951\\
   L3P3 &122 & 0.00493000 & 3846.87296 &  0.0057 & -0.0044 & 5.16 & 1070.14088 & 2.59473508\\
   L3P3 &121 & 0.00493000 & 3846.87940 &  0.0040 & -0.0054 &16.28 & 1070.14088 & 2.59474110\\
   L2R2 &121 & 0.00360000 & 3879.52758 &  0.0031 & -0.0030 & 8.81 & 1079.22542 & 2.59473332\\
   L2R2 &120 & 0.00360000 & 3879.53298 &  0.0021 & -0.0022 & 7.05 & 1079.22542 & 2.59473832\\
   L1P2 &119 &-0.00475000 & 3941.40916 &  0.0044 & -0.0034 &20.26 & 1096.43894 & 2.59473657\\
   L1R3 &119 &-0.00509000 & 3942.44064 &  0.0034 & -0.0042 & 8.44 & 1096.72534 & 2.59473835\\
   L1P2 &118 &-0.00475000 & 3941.41164 &  0.0031 & -0.0043 & 8.44 & 1096.43894 & 2.59473884\\
   L1P3 &118 &-0.00775000 & 3953.45055 &  0.0061 & -0.0050 & 8.44 & 1099.78718 & 2.59474144\\
   L0R0 &117 &-0.00800000 & 3983.42760 &  0.0040 & -0.0090 & 8.44 & 1108.12733 & 2.59473816\\
   L0P2 &117 &-0.01191000 & 3999.12069 &  0.0028 & -0.0028 & 8.44 & 1112.49600 & 2.59472815\\
   L0R3 &117 &-0.01202000 & 3999.42503 &  0.0084 & -0.0105 & 8.44 & 1112.58000 & 2.59473029\\
\enddata
\tablenotetext{a}{Transitions are labeled with L or W for Lyman or
Werner then the vibrational quantum number of the upper state, R,
Q or P transitions and finally the rotational quantum number of
the lower state}
\tablenotetext{b}{Vacuum wavelength}
\tablenotetext{c}{Barycentric redshift}
\end{deluxetable}

\clearpage

\clearpage

\clearpage

\clearpage

\end{document}